\pdfoutput=1

\documentclass{iopart}
\usepackage{iopams,array,booktabs,cite,dsfont,graphicx,multirow,setstack,subeqn,verbatim}
\usepackage[bookmarksnumbered,bookmarksopen=true,bookmarksopenlevel=1,linktocpage,pdfstartview=FitH]{hyperref}
\usepackage[all]{hypcap}

\newcolumntype{M}[1]{>{$}{#1}<{$}}
\newcolumntype{D}[1]{>{$\displaystyle}{#1}<{$}}
\newcolumntype{T}[1]{>{\begingroup\tiny$}{#1}<{$\endgroup}}
\newcolumntype{F}[1]{>{\begingroup\footnotesize$}{#1}<{$\endgroup}}

\newcommand{\half}{\case{1}{2}}
\newcommand{\fld}[1]{\ensuremath{\mathds{#1}}}
\newcommand{\rng}[1]{\ensuremath{\mathds{#1}}}
\newcommand{\rep}[1]{\ensuremath{\mathbf{#1}}}
\newcommand{\opname}[1]{\ensuremath{\mathrm{#1}}}

\newcommand{\mj}{\mathfrak{M}(\mathfrak{J}^{\fld{O}^s}_{3})}
\newcommand{\Js}{\mathfrak{J}^{\fld{O}^s}_{2}}
\newcommand{\J}{\mathfrak{J}^{\fld{O}^s}_{3}}
\newcommand{\mjz}{\mathfrak{M}(\mathfrak{J}^{\fld{O}_{\rng{Z}}^s}_{3})}
\newcommand{\jzs}{\mathfrak{J}_{2}^{\fld{O}_{\rng{Z}}^s}}
\newcommand{\jz}{\mathfrak{J}_{3}^{\fld{O}_{\rng{Z}}^s}}
\newcommand{\alg}{\mathds{A}}

\newcommand{\sO}{\fld{O}^s}

\newcommand{\sOz}{\fld{O}^{s}_{\rng{Z}}}
\newcommand{\sQz}{\fld{H}^{s}_{\rng{Z}}}
\newcommand{\emphlarge}{``large''\ }
\newcommand{\emphsmall}{``small''\ }

\newcommand{\SUSY}{\ensuremath{\mathcal{N}}}
\newcommand{\maxsymmsubsetneq}{\subsetneq_{\mathrm{symm}}^{\max}}
\newcommand{\orbitfour}[4]{\mathcal{O}_{\mathrm{1/#1-BPS, #2, }p=#3, D=#4}}
\newcommand{\orbittwo}[2]{\mathcal{O}_{\mathrm{1/#1-BPS, }D=#2}}
\newcommand{\modulifour}[4]{\mathcal{M}_{\mathrm{1/#1-BPS, #2, }p=#3, D=#4}}
\newcommand{\modulitwo}[2]{\mathcal{M}_{\mathrm{1/#1-BPS, }D=#2}}

\begin{document}

\article{\footnotesize{Imperial/TP/2010/mjd/1, CERN-PH-TH/2010-040, SU-ITP-10/07}}{Observations on Integral and Continuous U-duality Orbits in $\SUSY=8$ Supergravity}

\author{L Borsten$^1$, D Dahanayake$^1$, M J Duff$^1$, S Ferrara$^{2,3}$, {A Marrani}$^4$ and W Rubens$^1$}
\address{$^1$ Theoretical Physics, Blackett Laboratory, Imperial College London, London SW7 2AZ, United Kingdom}
\address{$^2$ Physics Department, Theory Unit, CERN, CH 1211, Geneva 23, Switzerland}
\address{$^3$ INFN - Laboratori Nazionali di Frascati, Via Enrico Fermi 40, 00044 Frascati, Italy}
\address{$^4$ Stanford Institute for Theoretical Physics, Stanford University, Stanford, CA 94305-4060, USA}
\eads{\href{mailto:leron.borsten@imperial.ac.uk}{\texttt{\textbf{leron.borsten@imperial.ac.uk}}}, \href{mailto:duminda.dahanayake@imperial.ac.uk}{\texttt{\textbf{duminda.dahanayake@imperial.ac.uk}}}, \href{mailto:m.duff@imperial.ac.uk}{\texttt{\textbf{m.duff@imperial.ac.uk}}}, \href{mailto:sergio.ferrara@cern.ch}{\texttt{\textbf{sergio.ferrara@cern.ch}}}, \href{mailto:marrani@lnf.infn.it}{\texttt{\textbf{marrani@lnf.infn.it}}}, \href{mailto:william.rubens06@imperial.ac.uk}{\texttt{\textbf{william.rubens06@imperial.ac.uk}}}}

\begin{abstract}
One would often like to know when two \emph{a priori} distinct extremal black $p$-brane solutions are in fact related by U-duality. In the classical supergravity limit the answer for a large class of theories has been known for some time now. However, in the full quantum theory the U-duality group is broken to a discrete subgroup, a consequence of the Dirac-Zwanziger-Schwinger charge quantization conditions. The question of U-duality orbits in this case is a nuanced matter.  In the present work we address this issue in the context of $\SUSY=8$ supergravity in four, five and six dimensions. The purpose of this note is to present and clarify what is currently known about these orbits while at the same time filling in some of the details not yet appearing in the literature. For the continuous case we present the cascade of relationships existing between the orbits, generated as one descends from six to four dimensions, together with the corresponding implications for the associated moduli spaces.  In addressing the discrete case we exploit the mathematical framework of \emph{integral Jordan algebras}, the \emph{integral Freudenthal triple system} and, in particular, the work of Krutelevich. The charge vector of the dyonic black string in $D=6$ is $SO(5,5; \rng{Z})$ related to a two-charge reduced canonical form uniquely specified by a set of two arithmetic U-duality invariants. Similarly, the black hole (string) charge vectors in $D=5$ are $E_{6(6)}(\rng{Z})$ equivalent to a three-charge canonical form, again uniquely fixed by a set of three arithmetic U-duality invariants.  However, the situation in four dimensions is, perhaps predictably, less clear. While black holes preserving more than 1/8 of the supersymmetries may be fully classified by known arithmetic $E_{7(7)}(\rng{Z})$ invariants, 1/8-BPS and non-BPS black holes yield increasingly subtle orbit structures, which remain to be properly understood. However, for the very special subclass of \emph{projective} black holes a complete classification is known. All projective black holes are $E_{7(7)}(\rng{Z})$ related to a four or five charge canonical form determined uniquely by the set of known arithmetic U-duality invariants. Moreover, $E_{7(7)}(\rng{Z})$ acts transitively on the charge vectors of projective black holes  with a given leading-order entropy.
\end{abstract}
\submitto{\CQG}
\maketitle

\section{Introduction}

The extremal black $p$-brane solutions of supergravity have played, and continue to play, a key role in unravelling the non-perturbative aspects of M-theory. Evidently, understanding  the structure of these solutions is of utmost importance. In particular, one would like to know how such solutions are interrelated by the set of global symmetries collectively known as U-duality. The electric/magnetic charge vectors of the asymptotically flat $p$-brane solutions form irreducible U-duality representations as in \autoref{tab:Mcharges}.
\begin{table}[!ht]
\caption{Asymptotically flat $p$-brane U-duality representations\label{tab:Mcharges}}
\begin{tabular*}{\textwidth}{@{\extracolsep{\fill}}c*{8}{M{c}}}
\toprule
$D$ & G                           & 0            & 1           & 2           & 3            & 4       & 5       & 6       \\
\midrule
10A & \fld{R}^{+}                    & \rep{1}      & \rep{1}     & \rep{1}     &              & \rep{1} & \rep{1} & \rep{1} \\
10B & SL(2,\fld{R})                  &              & \rep{2}     &             & \rep{1}      &         & \rep{2} &         \\
9   & SL(2,\fld{R})\times\fld{R}^{+}    & \rep{2+1}    & \rep{2}     & \rep{1}     & \rep{1}      & \rep{2} & \rep{2+1} \\
8   & SL(3,\fld{R})\times SL(2,\fld{R}) & \rep{(3',2)} & \rep{(3,1)} & \rep{(1,2)} & \rep{(3',1)} & \rep{(3,2)} \\
7   & SL(5,\fld{R})                  & \rep{10'}    & \rep{5}     & \rep{5'}    & \rep{10} \\
6   & SO(5,5;\fld{R})                & \rep{16}     & \rep{10}    & \rep{16'} \\
5   & E_{6(6)}(\fld{R})              & \rep{27}     & \rep{27'} \\
4   & E_{7(7)}(\fld{R})              & \rep{56} \\
\bottomrule
\end{tabular*}
\end{table}
In many relevant cases the macroscopic leading-order black $p$-brane entropy is a function of  these charges only, a result of the attractor mechanism \cite{Ferrara:1995ih,Strominger:1996kf,Ferrara:1996dd,Ferrara:1996um}. Consequently, an important question is whether two \emph{a priori} distinct black $p$-brane charge configurations are in fact related by U-duality. Mathematically this amounts to determining the distinct charge vector orbits under U-duality. In the classical limit the answer for a large class of theories has been known for some time now \cite{Ferrara:1997uz,Lu:1997bg,Ferrara:2006xx,Bellucci:2006xz}. In particular, for the maximally supersymmetric theories, obtained by the toroidal compactification of $D=11, \SUSY=1$ supergravity, a complete classification of all orbits in all dimensions $D\geq 4$ is known \cite{Ferrara:1997uz, Lu:1997bg}. However, in the full quantum theory the U-duality group is broken to a discrete subgroup, a consequence of the Dirac-Zwanziger-Schwinger charge quantization conditions \cite{Hull:1994ys}. Consequently, the U-duality orbits are furnished with a further level structural complexity, which, in some cases, is of particular mathematical significance \cite{Bhargava:2004,Krutelevich:2004}. However, the question of discrete U-duality orbits is not only interesting in its own right, it is also of physical importance with implications for a number of topics including the stringy origins of microscopic black hole entropy \cite{Maldacena:1999bp, Dabholkar:2007vk,Sen:2007qy,Banerjee:2007sr,Banerjee:2008pu,Banerjee:2008ri,Sen:2008ta,Sen:2008sp,Bianchi:2009mj}. Moreover, following a  conjecture of finiteness of $D=4, \SUSY=8$ supergravity \cite{Bern:2006kd}, it has recently been observed that some of the orbits of  $E_{7(7)}(\rng{Z})$ should play an important role in counting microstates of this theory \cite{Bianchi:2009mj}, even if it may differ from its superstring or M-theory completion \cite{Green:2007zzb}.

In the present work we address this issue in the context of $\SUSY=8$ supergravity in four, five and six dimensions. To this end we exploit the mathematical framework of \emph{integral Jordan algebras} and the \emph{integral Freudenthal triple system}, both of which have at their basis the ring of \emph{integral split-octonions} \cite{Gross:1996,Elkies:1996,Krutelevich:2002,Krutelevich:2004}. To a large extent this work is a continuation of the analysis used in studying the recently introduced black hole \emph{Freudenthal duality} \cite{Borsten:2009zy}, which in turn has its provenance in  recently established connections relating black hole entropy in M-theory to entanglement in quantum information theory \cite{Duff:2006uz,Kallosh:2006zs,Levay:2006kf, Duff:2006ue,Levay:2006pt,Duff:2006rf, Duff:2007wa,Bellucci:2007zi, Levay:2007nm, Borsten:2008ur,Levay:2008mi, Borsten:2008wd, Borsten:2009zy, Borsten:2009yb, Levay:2009bp,Borsten:2009ae, Rubens:2009zz}.

It is well known that the black holes and  strings appearing in the maximally supersymmetric 6-, 5- and 4-dimensional classical theories are elegantly described  by the exceptional  Jordan algebras and the closely related Freudenthal triple system (FTS) \cite{Gunaydin:1984ak,Gunaydin:1983bi,Gunaydin:1983rk,Ferrara:1997uz}.

In particular, the black string charge vectors of $D=6, \SUSY=8$ supergravity may be represented as elements of the exceptional Jordan algebra of $2\times 2$ Hermitian matrices defined over the split-octonions, denoted $\Js$. See \ref{sec:22J} for details. The \emph{reduced structure} group $\opname{Str}_0(\Js)$, defined as the set of invertible linear transformations preserving $N_2$, the quadratic norm \eref{eq:quadnorm}, is the $D=6$ U-duality group $SO(5,5;\fld{R})$,  under which the black string charges transform as the vector \rep{10}.  Moreover, in this case the quadratic norm $N_2$ is nothing but $I_2$, the  $SO(5,5;\fld{R})$ singlet in $\mathbf{10}\times\mathbf{10}$, which determines the black string entropy
\begin{equation*}
S_{D=6, \mathrm{BS}}\sim |I_2|.
\end{equation*}
See \emph{e.g.} \cite{Ferrara:1997uz,Andrianopoli:2007kz,Ferrara:2008xz}, and Refs. therein. There are two distinct charge vector orbits under $SO(5,5;\fld{R})$, one consisting of 1/2-BPS states and one consisting of 1/4-BPS states, distinguished respectively by the vanishing or not of  $I_2$ \cite{Lu:1997bg,Ferrara:2008xz}. Equivalently, these orbits may be distinguished by the \emph{rank} of the Jordan algebra element representing the charge vector, rank 1 states being 1/2-BPS while rank 2 states are 1/4-BPS. See \autoref{sec:D6cont} for details.

Similarly, the black hole charge vectors of $D=5, \SUSY=8$ supergravity may be represented as elements of the exceptional Jordan algebra of $3\times 3$ Hermitian matrices defined over the split-octonions, denoted $\J$  \cite{Ferrara:1997uz}. See \ref{sec:33J} for details. The reduced structure group $\opname{Str}_0(\J)$, defined as the set of invertible  linear transformations preserving $N_3$, the cubic norm \eref{eq:cubicnorm}, and the symmetric bilinear trace form \eref{eq:tracebilinearform}, is the $D=5$ U-duality group $E_{6(6)}(\fld{R})$,  under which the black hole charges transform as the fundamental \rep{27}.  Moreover, in this case the cubic norm $N_3$ is in fact $I_3$, the  $E_{6(6)}(\fld{R})$ singlet in $\mathbf{27}\times\mathbf{27}\times\mathbf{27}$, which determines the black hole entropy
\begin{equation*}
S_{D=5, \mathrm{BH}}=\pi\sqrt{ |I_3|}.
\end{equation*}
There are three distinct charge vector orbits under $E_{6(6)}(\fld{R})$, one 1/2-BPS, one 1/4-BPS and one 1/8-BPS,  distinguished by the vanishing or not of  $I_3$ and its derivatives \cite{Ferrara:1997ci}. See \autoref{sec:D5cont} for details. Again, these orbits may also be distinguished by the rank of the Jordan algebra element representing the charge vector. Rank 1 states are 1/2-BPS, rank 2 states are 1/4-BPS while  rank 3 states are 1/8-BPS \cite{Borsten:2008wd, Borsten:2009zy}. A directly analogous treatment goes through for the black string charges in $D=5$, which transform as the contragredient $\rep{27'}$ of $E_{6(6)}(\fld{R})$.

Finally, the black hole charge vectors of $D=4, \SUSY=8$ supergravity may be represented as elements of the Freudenthal triple system  denoted $\mj$. See \ref{sec:fts} for details. The \emph{automorphism} group $\opname{Aut}(\mj)$, defined as the set of invertible linear transformations preserving $\Delta$, the quartic norm \eref{eq:quarticnorm}, and the antisymmetric bilinear form \eref{eq:bilinearform}, is nothing but $E_{7(7)}(\fld{R})$, the $D=4$ U-duality group under which the black hole charges transform as the fundamental \rep{56}.  Moreover, the quartic norm $\Delta$ is exactly $I_4$, the  unique $E_{7(7)}(\fld{R})$ quartic invariant, which again determines the black hole entropy \cite{Kallosh:1996uy}
\begin{equation*}
S_{D=4, \mathrm{BH}}=\pi\sqrt{|I_4|}.
\end{equation*}
This is the first example exhibiting a non-BPS orbit. In total there are five distinct charge vector orbits under $E_{7(7)}(\fld{R})$, three of which have vanishing $I_4$, one 1/2-BPS, one 1/4-BPS and one 1/8-BPS, which are distinguished by the vanishing or not of the derivatives of $I_4$ \cite{Ferrara:1997ci}. The two orbits with non-vanishing $I_4$ are either 1/8-BPS or non-BPS according to whether $I_4>0$ or $I_4<0$ respectively. Again, these orbits may also be distinguished by the rank of the FTS element representing the charge vector. States of rank 1, 2 and 3 are 1/2-BPS, 1/4-BPS and 1/8-BPS respectively, all with vanishing $I_4$. Rank 4 states are split into 1/8-BPS and non-BPS as determined by the sign of $I_4$. See \autoref{sec:D4cont} for details. For further details concerning these orbits and their defining U-invariant BPS conditions the reader is referred to the original works \cite{Ferrara:1997ci, Ferrara:1997uz, Lu:1997bg}.

As one descends from $D=6$ to $D=4$, via spacelike dimensional reductions, a series of relationships connecting these U-duality orbits is generated. The  U-invariant BPS conditions in $D$ dimensions are ``embedded'' in those of $D-1$ dimensions, as is best understood by decomposing the $(D-1)$-dimensional U-duality group with respect to the $D$-dimensional U-duality group.  Taking care of whether or not the of charges of Kaluza-Klein vector are vanishing, one is then able to understand  how the various black $p$-brane solutions, their orbits and the associated moduli spaces are embedded under these spacelike reductions.  Moreover, in this way we can also study the reverse situation and so understand which higher dimensional black $p'$-brane solutions a given black $p$-brane in a given dimension may be ``uplifted'' to. It should be pointed out that the vanishing or not of the Kaluza-Klein vector charges is crucial in discriminating the various possible uplifts. A general result holding throughout the present treatment in $D =4, 5, 6$ can be stated as follows: if the charges of the Kaluza-Klein vector are not switched on, the supersymmetry preserving features of the solution are unaffected by the dimensional reduction. These results are presented  for $D=6\leftrightarrow D=5$ and $D=5\leftrightarrow D=4$ in \autoref{sec:5/6-Rels} and  \autoref{sec:4/5-Rels} respectively.

This summarizes the classification of the U-duality orbits for real valued charges.  However, as previously emphasized, the charges are actually quantized and the U-dualities are correspondingly broken to discrete subgroups as described in \cite{Hull:1994ys}.   The integral charge vector orbits are a nuanced matter and a complete characterization is as yet not known. Despite these additional complications, the discrete orbit classification is made possible in certain cases by the introduction of new \emph{arithmetic} U-duality invariants not appearing in the continuous case. These are typically given by the \emph{greatest common divisor} (gcd) of U-duality representations built out of the basic charge vector representations \cite{Krutelevich:2002,Krutelevich:2004,Dabholkar:2007vk,Sen:2007qy,Banerjee:2007sr,Banerjee:2008pu,Banerjee:2008ri,Sen:2008ta,Sen:2008sp, Borsten:2009zy}. One purpose of this note is to present and clarify what is currently known about these discrete orbits while at the same time filling in some of the details not yet appearing in the literature.

Let us now briefly summarize the present situation. An important general observation is that, since the conditions separating the continuous orbits are manifestly invariant under the corresponding discrete U-dualities, those states unrelated in the continuous cases remain unrelated in the discrete case. Consequently, the discrete orbits fall into disjoint sets corresponding directly to the orbits of the classical theory. A second important observation, emphasized in \cite{Ferrara:2008xz}, is that the gcd of a U-duality representation, built out of the relevant basic charge vector representation, is only well defined if that representation is non-vanishing. In practice this means first computing which class of orbits as defined by the continuous analysis a given state lies in. This, in turn, determines the subset of the arithmetic invariants that are well defined for this particular state. It is this subset that is then to be used in specifying the particular discrete orbit to which the state belongs, the remaining arithmetic invariants being ill-defined and contentless.

Beginning in $D=6$ the integral charge vectors of the dyonic black strings may be represented as elements of the integral Jordan algebra of $2\times 2$ Hermitian matrices defined over the ring of \emph{integral} split-octonions, denoted $\jzs$. See \ref{sec:22Jz} for details. The discrete U-duality group $SO(5,5; \rng{Z})$ is given by the set of invertible $\rng{Z}$-linear transformations preserving the quadratic norm \eref{eq:quadnorm}. An arbitrary charge vector is $SO(5,5; \rng{Z})$  related to a two-charge reduced canonical form \eref{eq:6dcan} uniquely specified by a set of two arithmetic U-duality invariants \eref{eq:6dinvs}. The two orbits of the continuous case now form two disjoint countably infinite sets of discrete orbits which may be parametrized using the arithmetic invariants.

Similarly, the black hole  charge vectors in $D=5$ may be represented as elements of the integral Jordan algebra of $3\times 3$ Hermitian matrices defined over the ring of integral split-octonions, denoted $\jz$. See \ref{sec:33Jz} for details. The discrete U-duality group $E_{6(6)}(\rng{Z})$ is given by the set of invertible $\rng{Z}$-linear transformations preserving the cubic norm \eref{eq:cubicnorm} and the trace bilinear form \eref{eq:tracebilinearform}. An arbitrary charge vector is $E_{6(6)}(\rng{Z})$ equivalent to a three-charge canonical form \eref{eq:5dcan}, again uniquely fixed by a set of three arithmetic U-duality invariants \eref{eq:5dinvs}. The three orbits of the continuous case now form three disjoint classes of discrete orbits which may be parametrized using the arithmetic invariants \cite{Krutelevich:2002, Borsten:2009zy}. A directly analogous treatment goes through for the black string charges in $D=5$, which transform as the contragredient $\rep{27'}$ of $E_{6(6)}(\rng{Z})$.

However, the situation in four dimensions is, perhaps predictably, less clear. The black hole  charge vectors  may be represented as elements of the integral FTS defined over $\jz$. See \ref{sec:fts} for details. The discrete U-duality group $E_{7(7)}(\rng{Z})$ is given by the set of invertible $\rng{Z}$-linear transformations preserving the quartic norm \eref{eq:quarticnorm} and the antisymmetric bilinear form  \eref{eq:bilinearform}. An arbitrary charge vector is $E_{7(7)}(\rng{Z})$ equivalent to a five-charge canonical form \eref{eq:4dcan}. However, this canonical form is not uniquely fixed by the known set of arithmetic U-duality invariants \eref{eq:4dinvs}. Despite this, for particular subcases more can be said. Indeed, the classes of discrete orbits corresponding to the 1/2-BPS and 1/4-BPS continuous orbits may be completely classified using the known arithmetic invariants \cite{Krutelevich:2004}, as described in \autoref{sec:D4disc}.  For those black hole preserving less than 1/4 of the supersymmetries the orbit structure becomes more complicated and the orbit classification is not known. However, even in this case a full classification is possible for the of \emph{projective} black holes. See \autoref{sec:proj} for details. The concept of projectivity was originally introduced in the in the number-theoretic context of \cite{Bhargava:2004} where such elements are mapped to invertible ideal classes of quadratic rings. This notion was later generalized by Krutelevich in \cite{Krutelevich:2004} with a view to understanding $E_{7(7)}(\rng{Z})$ orbit structure.  Indeed, $E_{7(7)}(\rng{Z})$ acts transitively on the set of projective black holes of a given quartic norm. Moreover, they are $E_{7(7)}(\rng{Z})$ equivalent to  a simplified four or five charge canonical form \eref{eq:projcan} depending on whether the quartic norm is even or odd respectively \cite{Krutelevich:2004, Borsten:2009zy}.

This note is organized as follows. In \autoref{sec:D6} we begin by recalling the continuous U-duality orbits in $D=6$, emphasizing the Jordan algebraic perspective, before presenting in detail the corresponding discrete treatment. In \autoref{sec:D5} the same treatment is applied to continuous U-duality black hole charge orbits in $D=5$. Subsequently, the intricate web of relations connecting the orbits and moduli spaces of the 6-dimensional theory to those of the 5-dimensional theory are presented. This analysis is concluded with corresponding discrete U-duality treatment of the $D=5$ integral black hole charge orbits. The same continuous analysis is undertaken for $D=4$ in \autoref{sec:D4}, completing the cascade of relationships between the orbits and moduli spaces in 6-, 5- and 4-dimensions. This is followed by the  discrete U-duality treatment of the integral black hole charge orbits.  We conclude with a summary of open questions. For the most part the technical details are relegated to the appendices in an effort to avoid an oppressive number of formal definitions in the main body of the text. In \ref{sec:octonions} we present the minimal background necessary to introduce the ring of integral split-octonions underlying this analysis. In \ref{sec:6DJ} we describe the continuous and integral quadratic Jordan algebras together with their application to black strings in $D=6$. In \ref{sec:5DJ} we describe the continuous and integral cubic Jordan algebras together with their application to black holes (strings) in $D=5$.  In \ref{sec:fts} we describe the continuous and integral FTS, defined over the $D=5$ Jordan algebra, together with its application to black holes in $D=4$.

\section{\texorpdfstring{Black strings in $D=6$}{Black strings in D=6}}\label{sec:D6}

\subsection{\texorpdfstring{U-duality orbits of $SO(5,5;\fld{R})$}{U-duality orbits of SO(5,5;R)}}\label{sec:D6cont}

In the classical supergravity limit the $5+5$ electric/magnetic black string charges form an $SO(5,5;\fld{R})$ vector  $\mathcal{Q}_{r}$ ($r=1,...,10$ throughout). Under  $SO(1,1;\fld{R})\times SO(4,4;\fld{R})$  the vector breaks as
\begin{equation}
\mathbf{10}\to
\mathbf{1}_{2}+\mathbf{1}_{-2}+\mathbf{8}_{v0}.
\end{equation}
where the singlets lie in the NS-NS sector and correspond to a fundamental string  and an NS5-brane, while the $\mathbf{8}_v$ is made up of R-R charges. In this basis the charges $\mathcal{Q}_{r}$ may be conveniently represented as an element $\mathcal{Q}$ of the Jordan algebra $\mathfrak{J}^{\fld{O}^s}_{2}$ of
\emph{split}-octonionic $2\times 2$ Hermitian matrices,
\begin{equation}
\mathcal{Q}=\left(\begin{array}{cc}p^0&Q_v\\\overline{Q}_v&q_0\end{array}\right),
\ \  \mathrm{where} \ \  q^0,p_0\in \fld{R} \ \  \mathrm{and}
\ \  Q_v\in\fld{O}^s.
\end{equation}
The set of linear invertible transformations leaving the quadratic norm,
\begin{equation}
N_2(\mathcal{Q})=\det(\mathcal{Q}),
\end{equation}
invariant is the $D=6$ U-duality group $SO(5,5;\fld{R})$. Using the dictionary
\eref{eq:dictD6} one finds,
\begin{equation}
N_2(\mathcal{Q})=I_2(\mathcal{Q}),
\end{equation}
where,
\begin{equation}\label{eq:I_2-el}
I_2(\mathcal{Q})=\eta^{rs}\mathcal{Q}_{r} \mathcal{Q}_{s}
\end{equation}
and $\eta^{rs}$ is the $SO(5,5;\fld{R})$ metric,
\begin{equation}\label{eq:I_2metric}
\eta=\left(\begin{array}{cc}0&\mathds{1}\\\mathds{1}&0\end{array}\right).
\end{equation}
The black string entropy is proportional to the quadratic norm,
\begin{equation}
S_{D=6, \mathrm{BS}}\sim |I_2(\mathcal{Q})|= |N_2(\mathcal{Q})|.
\end{equation}
See \emph{e.g.} \cite{Ferrara:1997uz,Andrianopoli:2007kz,Ferrara:2008xz}, and Refs. therein.

There are two U-duality orbits, one 1/2-BPS \emphsmall orbit and one 1/4-BPS \emphlarge orbit \cite{Lu:1997bg, Ferrara:1997ci, Ferrara:2008xz}. Note, an orbit is referred as \emphsmall  in the sense that the associated U-duality invariant and, hence, Bekenstein-Hawking entropy are vanishing. Correspondingly, for \emphlarge orbits the U-duality invariant and Bekenstein-Hawking entropy are non-vanishing. These orbits may distinguished by the Jordan rank
of $\mathcal{Q}$ as detailed in \ref{sec:quadJ},
\begin{equation}
\begin{tabular}{@{Rank }c@{\quad}M{c}@{\quad}c@{-BPS,}}
1 & \mathcal{Q}\not=0, N_2(\mathcal{Q})=0 & 1/2\\[3pt]
2 & N_2(\mathcal{Q})\not=0                & 1/4
\end{tabular}
\end{equation}
which is precisely the condition originally presented in \cite{Ferrara:1997ci}. The orbits, their rank conditions, dimensions and representative states are summarized in \autoref{tab:J2rank}.
\begin{table}
\caption{$D=6$ black string orbits, their corresponding rank conditions, dimensions and SUSY.\label{tab:J2rank}}
\begingroup
\footnotesize
\begin{tabular*}{\textwidth}{@{\extracolsep{\fill}}*{8}{D{c}}}
\toprule
\multirow{2}{*}{Rank} & \multicolumn{2}{c}{Rank/orbit conditions} & \multirow{2}{*}{Representative state} & \multirow{2}{*}{Orbit } & \multirow{2}{*}{dim} & \multirow{2}{*}{SUSY} \\
\cmidrule(r){2-3}
  & \mathrm{non\mbox{-}vanishing} & \mathrm{vanishing} &            &                                          &   &     \\
\midrule
1 & \mathcal{Q}          & N_2(\mathcal{Q}) & \opname{diag}(1,0) & \frac{SO(5,5;\fld{R})}{SO(4,4;\fld{R})\rtimes\fld{R}^8} & 9 & 1/2 \\
2 & N_2(\mathcal{Q})     & -                & \opname{diag}(1,k) & \frac{SO(5,5;\fld{R})}{SO(5,4;\fld{R})}            & 9 & 1/4 \\
\bottomrule
\end{tabular*}
\endgroup
\end{table}

\subsection{\texorpdfstring{U-duality orbits of $SO(5,5;\rng{Z})$}{U-duality orbits of SO(5,5;Z)}}\label{sec:D6disc}

For quantized charges the continuous  U-duality is broken to an infinite discrete subgroup, which for $D=6$ is given by $SO(5,5;\rng{Z})\subset SO(5,5;\fld{R})$ \cite{Hull:1994ys}. The integral Jordan algebra, $\mathfrak{J}^{\fld{O}_{\rng{Z}}^s}_{2}$, of integral split-octonionic $2\times 2$ Hermitian matrices provides a natural model for $SO(5,5;\rng{Z})$, which may used to analyse the discrete U-duality orbits. The quantized black string  charge vector is given by,
\begin{equation}
\mathcal{Q}=\left(\begin{array}{cc}p^0&Q_v\\\overline{Q}_v&q_0\end{array}\right),
\ \  \mathrm{where} \ \  q^0,p_0\in \rng{Z} \ \  \mathrm{and}
\ \  Q_v\in\fld{O}_{\rng{Z}}^s.
\end{equation}
The discrete group $SO(5,5;\rng{Z})$ is defined by the set of norm-preserving invertible $\rng{Z}$-linear transformations,
\begin{equation}
\{\sigma:\mathfrak{J}^{\fld{O}_{\rng{Z}}^s}_{2}\to\mathfrak{J}^{\fld{O}_{\rng{Z}}^s}_{2}|N_2(\sigma(\mathcal{Q}))=N_2(\mathcal{Q})\}.
\end{equation}
It is with this framework that we shall study the discrete U-duality orbits.

The first important observation is that the charge conditions defining the orbits in the continuous theory are manifestly invariant under the discrete subgroup $SO(5,5;\rng{Z})$ and, hence, those states unrelated by U-duality in the classical theory remain unrelated in the quantum theory.   There are two disjoint \emph{classes} of orbits, one 1/2-BPS and one 1/4-BPS, corresponding to the two orbits of the continuous case. However, each of these classes is broken up into a countably infinite set of discrete orbits. To classify these orbits we use $SO(5,5;\rng{Z})$ to bring an arbitrary charge vector into a \emph{diagonal reduced} canonical form, which is uniquely defined by the following set of two discrete invariants,
\begin{equation}\label{eq:6dinvs}
\begin{array}{r@{\ :=\ }l}
b_1(\mathcal{Q})&\gcd(\mathcal{Q}),\\
b_2(\mathcal{Q})&N_2(\mathcal{Q}).
\end{array}
\end{equation}
See \ref{sec:22Jz} for details.

\paragraph{$D=6$ diagonal reduced canonical form} Every element $\mathcal{Q}\in\mathfrak{J}^{\fld{O}^{s}_{\rng{Z}}}_{2}$ is $SO(5,5;\rng{Z})$ equivalent to a diagonally reduced canonical form,
\begin{equation}\label{eq:6dcan}
\mathcal{Q}_{\mathrm{can}}=k\left(\begin{array}{cc}1&0\\0&l\end{array}\right),\ \  \mathrm{where} \ \  k>0, |l|\geq 0.
\end{equation}
The canonical form is uniquely determined by \eref{eq:6dinvs} since
\begin{equation}
\begin{array}{r@{\ =\ }l}
b_1(\mathcal{Q}_{\mathrm{can}})&k,\\
b_2(\mathcal{Q}_{\mathrm{can}})&k^2l,
\end{array}
\end{equation}
so that for arbitrary $\mathcal{Q}$ one obtains $k=b_1(\mathcal{Q})$ and $l=k^{-2}b_2(\mathcal{Q})$.

\paragraph{$D=6$ Black string orbit classification}
\begin{enumerate}
\item The complete set of distinct 1/2-BPS charge vector orbits is given by,
\begin{equation}
\big\{\left(\begin{array}{cc}k&0\\0&0\end{array}\right), \ \  \mathrm{where}\ \  k>0 \big\}.
\end{equation}

\item The complete set of distinct 1/4-BPS charge vector orbits is given by,
\begin{equation}
\big\{\left(\begin{array}{cc}k&0\\0&kl\end{array}\right), \ \  \mathrm{where}\ \  k,|l|>0 \big\}.
\end{equation}

\end{enumerate}

\section{\texorpdfstring{Black holes in $D=5$}{Black holes in D=5}}\label{sec:D5}

\subsection{\texorpdfstring{U-duality orbits of $E_{6(6)}(\fld{R})$}{U-duality orbits of E\_6(6)(R)}}\label{sec:D5cont}

In the classical supergravity limit the 27 $Q_i$  ($i=1,\ldots , 27$ throughout) electric  black hole  charges transform as the fundamental \rep{27} of the continuous U-duality group $E_{6(6)}(\fld{R})$. Under $SO(1,1;\fld{R})\times SO(5,5;\fld{R})$ the \rep{27} breaks as
\begin{equation}\label{eq:27-branching}
\rep{27}\to\rep{1}_{4}+\rep{10}_{-2}+\rep{16}_{1}
\end{equation}
where the singlet may be identified as the graviphoton charge descending from $D=6$, the \rep{10} as the remaining NS-NS sector charges and the \rep{16} as the R-R sector charges. Further decomposing under $SO(4,4;\fld{R})$ one obtains
\begin{equation}
\rep{27}\to\rep{1}+\rep{1}+\rep{1}+\rep{8}_{v}+\rep{8}_{s}+\rep{8}_{c}.
\end{equation}
In this basis the charges $Q_i$ may be conveniently represented as an element $Q$ of the cubic Jordan algebra $\mathfrak{J}^{\fld{O}^s}_{3}$ of \emph{split}-octonionic $3\times 3$ Hermitian matrices,
\begin{equation}\fl\qquad
Q=\left(\begin{array}{ccc}q_1&Q_s&\overline{Q_c}\\\overline{Q_s}&q_2&Q_v\\Q_c&\overline{Q}_v&q_3\end{array}\right), \ \  \mathrm{where} \ \  q_1,q_2, q_3 \in \fld{R} \ \  \mathrm{and} \ \  Q_{v,s,c}\in\fld{O}^s.
\end{equation}
The cubic norm $N_3$ \eref{eq:cubicnormJ33} is then given by the determinant like object,
\begin{equation}
\fl N_3(Q)=q_1q_2q_3-q_1 Q_v\overline{Q_v}-q_2 Q_c\overline{Q_c}-q_3 Q_s \overline{Q_s} +(Q_vQ_c)Q_s+\overline{Q_s}(\overline{Q_c}\overline{Q_v}).
\end{equation}
The set of invertible linear transformations leaving the cubic norm and trace bilinear form invariant is nothing but the $D=5$ U-duality group $E_{6(6)}(\fld{R})$. Moreover,
\begin{equation}\label{eq:I_3-el}
N_3(Q)=I_3(Q),
\end{equation}
where,
\begin{equation}
I_3(Q)=\frac{1}{3!}d^{ijk}Q_i Q_j Q_k
\end{equation}
and $d^{ijk}$ is the $E_{6(6)}(\fld{R})$ invariant tensor.

The black hole entropy is simply given by the cubic norm,
\begin{equation}\label{eq:D=5-entropy}
S_{\mathrm{BS}}=\pi\sqrt{|N_3(Q)|}=\pi\sqrt{|I_3(Q)|}.
\end{equation}

In this case there are three U-duality orbits,  1/2-BPS and  1/4-BPS \emphsmall orbits and a single 1/8-BPS \emphlarge orbit \cite{Ferrara:1997uz}. These orbits may distinguished by the Jordan rank of $Q$,
\begin{equation}
\begin{tabular}{@{Rank }c@{\quad}M{c}@{\quad}c@{-BPS,}}
1& Q\not=0, Q^\sharp=0      & 1/2\\[3pt]
2& Q^\sharp\not=0, N_3(Q)=0 & 1/4\\[3pt]
3& N_3(Q)\not=0             & 1/8
\end{tabular}
\end{equation}
where $^\sharp$ is the quadratic adjoint map \eref{eq:quadadjexp}.  See \ref{sec:cubicJ} for details. Note, $Q^\sharp$ transforms as a $\mathbf{27}'$ under $E_{6(6)}(\fld{R})$. Similarly
\begin{equation}
\partial_i I_3(Q)=\frac{\partial I_3(Q)}{\partial Q_i}\sim d^{ijk}Q_jQ_k
\end{equation}
manifestly transforms as a $\mathbf{27}'$ under $E_{6(6)}(\fld{R})$ so that $Q^\sharp\sim\partial_i I_3(Q)$. Hence, these conditions are entirely equivalent to the conditions originally presented in
\cite{Ferrara:1997ci},
\begin{equation}
\begin{tabular}{@{Rank }c@{\quad}M{c}@{\quad}c@{-BPS}l}
1 & Q\not=0, \partial_i I_3(Q)=0      & 1/2&,\\[3pt]
2 & \partial_i I_3(Q)\not=0, I_3(Q)=0 & 1/4&,\\[3pt]
3 & I_3(Q)\not=0                      & 1/8&.
\end{tabular}
\end{equation}
The orbits with their rank conditions, dimensions and representative states are summarized in \autoref{tab:Jrank}.
The 27 magnetic black string charges $P^i$ form the contragredient $\rep{27'}$ of $E_{6(6)}(\fld{R})$. The orbit classification is identical to the black hole case.
\begin{table}
\caption{$D=5$ black hole orbits, their corresponding rank conditions, dimensions and SUSY.\label{tab:Jrank}}
\begingroup
\footnotesize
\begin{tabular*}{\textwidth}{@{\extracolsep{\fill}}*{7}{D{c}}}
\toprule
\multirow{2}{*}{Rank} & \multicolumn{2}{c}{Rank/orbit conditions} &\multirow{2}{*}{Representative state} & \multirow{2}{*}{Orbit} &\multirow{2}{*}{dim} & \multirow{2}{*}{SUSY} \\
\cmidrule(r){2-3}
& \mathrm{non\mbox{-}vanishing} & \mathrm{vanishing} \\
\midrule
1 & Q        & Q^\sharp & \opname{diag}(1,0,0) & \frac{E_{6(6)}(\fld{R})}{O(5,5;\fld{R})\ltimes\fld{R}^{16}} & 17 & 1/2 \\
2 & Q^\sharp & N_3(Q)   & \opname{diag}(1,1,0) & \frac{E_{6(6)}(\fld{R})}{O(5,4;\fld{R})\ltimes\fld{R}^{16}} & 26 & 1/4 \\
3 & N_3(Q)   & -        & \opname{diag}(1,1,k) & \frac{E_{6(6)}(\fld{R})}{F_{4(4)}(\fld{R})}            & 26 & 1/8 \\
\bottomrule
\end{tabular*}
\endgroup
\end{table}

\subsection{\texorpdfstring{$D=5 \longleftrightarrow D=6$ relations for charge orbits and moduli spaces}{D=5 <-> D=6 relations for charge orbits and moduli spaces}}\label{sec:5/6-Rels}

Through the branching of the $\mathbf{27}^{(\prime }{}^{)}$ irrep. of $D=5$ U-duality $E_{6(6)}(\fld{R})$ with respect to $D=6$ $U$-duality $SO(5,5;\fld{R}) \left( \times SO\left( 1,1;\fld{R}\right) \right) $ \cite{Ferrara:1997ci,Andrianopoli:2007kz,Bianchi:2009mj} ($r=1,...,10$, and $\alpha =1,...,16$ throughout; cf. Eq. \eref{eq:27-branching})
\begin{subequations}
\begin{eqnarray}
&\left\{\begin{array}{cl}\rep{27}&\to\rep{1}_4+\rep{10}_{-2}+\rep{16}_1,\\Q_i&\to\left( q_{z},\mathcal{Q}_{r},q_{\alpha }\right);\end{array}\right.\label{eq:br-1}\\
&\left\{\begin{array}{cl}\rep{27'}&\to\rep{1}_{-4}+\rep{10}_2+\rep{16'}_{-1},\\P^i&\to \left( p^{z},\mathcal{P}^{r},p^{\alpha }\right),\end{array}\right.
\end{eqnarray}
\end{subequations}
the unique electric and magnetic invariants \eref{eq:I3-el} and \eref{eq:I3-magn} of the \rep{27} can be written in a manifestly $SO(5,5; \fld{R}) $-invariant way respectively as follows \cite{Ferrara:1997ci,Ferrara:2006yb}:
\begin{subequations}
\begin{eqnarray}
I_{3}\left( Q\right)  &=\frac{1}{2}\left[ q_{z}I_{2}\left( \mathcal{Q}\right) +\mathcal{Q}_{r}\left( \gamma ^{r}\right) ^{\alpha \beta}q_{\alpha}q_{\beta }\right];\label{eq:I3-I2-el} \\
I_{3}\left( P\right)  &=\frac{1}{2}\left[ p^{z}I_{2}\left( \mathcal{P}\right) +\mathcal{P}^{r}\left( \gamma ^{r}\right) _{\alpha \beta}p^{\alpha }p^{\beta }\right],\label{eq:I3-I2-magn}
\end{eqnarray}
\end{subequations}
where $\gamma ^{r}$'s are the $SO(5,5; \fld{R})$-gamma matrices, $q_{z}$ and $p^{z}$ respectively are the electric and magnetic charge of the $D=6\to 5$ Kaluza-Klein vector, and the unique quadratic invariant of the \rep{10} of $SO(5,5; \fld{R})$ is defined as follows (note the basis change compared to \eref{eq:I_2-el}-\eref{eq:I_2metric} $\eta _{rs}=\opname{diag}\left(1,\ldots,1,-1,\ldots,-1\right) $):
\begin{subequations}
\begin{eqnarray}
I_{2}\left( \mathcal{Q}\right)  &:=\eta^{rs}\mathcal{Q}_{r}\mathcal{Q}_{s};\\
I_{2}\left( \mathcal{P}\right)  &:=\eta_{rs}\mathcal{P}^{r}\mathcal{P}^{s}.
\end{eqnarray}
\end{subequations}

In order to study the relations among the various charge orbits of maximal supergravity in $D=5$ and $D=6$ (and the consequences for the related moduli spaces\footnote{In all the treatment $M$, $\mathcal{O}$ and $\mathcal{M}$ respectively denote a scalar manifold, a charge orbit and a \emph{moduli space}. $\mathcal{M}$ is defined all along the scalar flow, from the near horizon geometry (\emph{if any}, at the classical level) to asymptotically flat spatial infinity. Thus, if the corresponding $\mathcal{O}$ is \emphlarge, $\mathcal{M}$ can be interpreted as the \emph{moduli space} of attractor solutions (at the near horizon geometry) \emph{and} the moduli space of the ADM mass (at spatial infinity). In the case of \emphsmall $\mathcal{O}$, the attractor near horizon interpretation of corresponding $\mathcal{M}$ breaks down.}), let us briefly recall the U-invariant classification of the charge orbits of black $p=0,2$-branes (black holes, respectively black membranes) in $\SUSY=\left(2,2\right) $, $D=6$ supergravity \cite{Lu:1997bg,Ferrara:1997ci}.

\subsubsection{\texorpdfstring{R\'{e}sum\'{e} on $p=0,2$ black branes in $\SUSY=\left( 2,2\right) $, $D=6$}{R\'{e}sum\'{e} on p=0,2 black branes in N=(2,2), D=6}}\label{sec:p=0,2}

Without any loss of generality, let us consider a black hole ($p=0$) corresponding to the $D=6$ uplift of a $D=5$ black hole. In other words, we only consider  the orbits of chiral spinor \rep{16} of $SO\left(5,5; \fld{R}\right) $ determined by the branching \eref{eq:br-1} of \rep{27} of $E_{6(6)}(\fld{R})$.

\begin{enumerate}
\item  1/4-BPS \emphsmall orbit ($2$-charges solution) \cite{Lu:1997bg,Ferrara:1997ci}
\begin{equation}\label{eq:PA-13}
\orbitfour{4}{small}{0,2}{6}=\frac{SO(5,5; \fld{R}) }{SO( 3,4;\fld{R}) \rtimes \fld{R}^{8}},
\end{equation}
defined by
\begin{equation}
\left( \gamma ^{r}\right) ^{\alpha \beta }q_{\alpha }q_{\beta }\neq0,
\end{equation}
for \emph{at least} some $r=1,...,10$. Namely, the $SO(5,5; \fld{R}) $-invariant constraint defining such an orbit is the fact that $q_{\alpha }$ is \emph{not} a \emph{pure} chiral spinor of $SO\left( 5,5;\fld{R}\right) $. The resulting Bekenstein-Hawking entropy is vanishing:
\begin{equation}\label{eq:PA-1}
S_{\mathrm{BH, }D=6}=0.
\end{equation}

\item  1/2-BPS \emphsmall orbit ($1$-charge solution) \cite{Lu:1997bg,Ferrara:1997ci}
\begin{equation}\label{eq:PA-14}
\orbitfour{2}{small}{0,2}{6}=\frac{SO(5,5; \fld{R}) }{SL\left( 5,\fld{R}\right) \rtimes \fld{R}^{10}},
\end{equation}
defined by
\begin{equation}
\left( \gamma ^{r}\right) ^{\alpha \beta }q_{\alpha }q_{\beta
}=0,~\forall r.
\end{equation}
In other words, the $SO(5,5; \fld{R}) $-invariant constraint defining such an orbit is the fact that $q_{\alpha }$ is a \emph{pure} chiral spinor of $SO\left( 5,5;\fld{R}\right) $. However, it is worth recalling that the non-triviality of the background implies that \emph{at least} some $\alpha \in \left\{ 1,...,16\right\} $ exist such that
\begin{equation}
\frac{\partial }{\partial q_{\delta }}\left[ \left( \gamma^{r}\right) ^{\alpha \beta }q_{\alpha }q_{\beta }\right] =2\left(\gamma ^{r}\right) ^{\delta \beta }q_{\beta }\neq 0.
\end{equation}
As for case $1$, the resulting Bekenstein-Hawking entropy is vanishing (see \eref{eq:PA-1}).
\end{enumerate}

\subsubsection{\texorpdfstring{The 1/4-BPS black string orbit under $D=6$ $\to 5$}{The 1/4-BPS black string orbit under D=6 -> 5}}\label{sec:p=1-1/4}

A representative of the 1/4-BPS \emphlarge charge orbit
\begin{equation}\label{eq:PA-11}
\orbitfour{4}{large}{1}{6}=\frac{SO(5,5; \fld{R}) }{SO\left( 5,4;\fld{R}\right) }
\end{equation}
of a dyonic black string in $\SUSY=\left( 2,2\right) $, $D=6$ supergravity is provided by
\begin{subequations}\label{eq:a1to6}
\begin{equation}\label{eq:a1to3}
\begin{array}{rl}
P^{i}           &=0;  \\
q_{\alpha }     &=0;  \\
\mathcal{Q}_{r} &:I_{2}\left( \mathcal{Q}\right) \neq 0.
\end{array}
\end{equation}
By plugging \eref{eq:a1to3} into \eref{eq:I3-I2-el}, one obtains
\begin{equation}\label{eq:a4}
I_{3}\left( Q\right) =\half q_{z}I_{2}\left( \mathcal{Q}\right).
\end{equation}

The treatment splits in two separate cases, depending on the vanishing or not of the magnetic charge of Kaluza-Klein vector:

\begin{enumerate}
\item  In the case
\begin{equation}
q_{z}=0,
\end{equation}
Eq. \eref{eq:a4} yields
\begin{equation}\label{eq:a5}
\left\{
\begin{array}{c@{\ =\ }l}
I_{3}\left( Q\right)&0; \\
\frac{\partial I_{3}\left( Q\right) }{\partial q_{z}}&\half I_{2}\left(\mathcal{Q}\right) \neq 0,
\end{array}
\right.
\end{equation}
corresponding to a 1/4-BPS \emphsmall black hole of $\SUSY=8$, $D=5$ supergravity.

\item  On the other hand, for
\begin{equation}
q_{z}\neq 0,
\end{equation}
then Eq. \eref{eq:a4} gives
\begin{equation}\label{eq:a6}
I_{3}\left( q\right) \neq 0,
\end{equation}
corresponds to a 1/8-BPS ``large'' black hole of $\SUSY=8$, $D=5$ supergravity.
\end{enumerate}
\end{subequations}

Thus, at the level of charge orbits, Eqs. \eref{eq:a1to6} correspond to the following picture:
\begin{equation}\label{eq:a7}\fl\qquad
\begin{array}{c}
\orbitfour{4}{large}{1}{6}=\frac{SO(5,5; \fld{R}) }{SO\left( 5,4;\fld{R}\right)} \\
\overset{\displaystyle\swarrow}{\orbittwo{4}{5}=\frac{E_{6\left( 6\right)}(\fld{R})}{SO\left(5,4;\fld{R}\right) \rtimes \fld{R}_{16}}} \qquad\qquad
\overset{\displaystyle\searrow}{\orbittwo{8}{5}=\frac{E_{6(6)}(\fld{R})}{F_{4\left( 4\right) }(\fld{R})}.}
\end{array}
\end{equation}
Indeed, at the level of the semi-simple part of the orbit stabilizers, the embedding\footnote{``max'' and ``symm'' respectively denote the maximality and symmetricity of the group embedding under consideration.}
\begin{equation}
SO\left( 5,4;\fld{R}\right) \maxsymmsubsetneq F_{4\left( 4\right) }(\fld{R})
\end{equation}
holds. At the level of corresponding moduli spaces, \eref{eq:a7} implies that (see also \eref{eq:PPPP-1})
\begin{equation}
\modulifour{4}{large}{1}{6}=\frac{SO(5,4;\fld{R})}{SO(5,\fld{R})\times SO(4,\fld{R})}
\end{equation}
satisfies
\begin{equation}\label{eq:emb-1}
\modulifour{4}{large}{1}{6}\subsetneq ~\modulitwo{8}{5},
\end{equation}
which is nothing but a part of the embedding \eref{eq:PPP-1}.

Thus, it follows that a \emphsmall 1/4-BPS black hole, as well as a \emphlarge 1/8-BPS black hole, of $\SUSY=8$, $D=5$ supergravity can be uplifted to a \emphlarge 1/4-BPS dyonic black string of $\SUSY=\left( 2,2\right) $, $D=6$ supergravity.

It should be pointed out that the non-vanishing or not of charges of Kaluza-Klein vector is crucial in order to discriminate among the various possible uplifts. A general result holding throughout the present treatment in $D=4,5,6$ can be stated as follows: if the charges of the Kaluza-Klein vector are \emph{not} switched on, the supersymmetry-preserving features of the solution are \emph{unaffected} by the dimensional reduction.

\subsubsection{\texorpdfstring{The 1/2-BPS black string orbit under $D=6$ $\to 5$}{The 1/2-BPS black string orbit under D=6 -> 5}}\label{sec:p=1-1/2}

A representative of the 1/2-BPS \emphsmall charge orbit
\begin{equation}\label{eq:PA-12}
\orbitfour{2}{small}{1}{6}=\frac{SO(5,5; \fld{R}) }{SO\left( 4,4;\fld{R}\right) \rtimes \fld{R}^{8}},
\end{equation}
of a dyonic black string in $\SUSY=\left( 2,2\right) $, $D=6$ supergravity is provided by
\begin{subequations}\label{eq:b1to6}
\begin{equation}\label{eq:b1to3}
\begin{array}{cl}
P^{i}       &=0;  \\
q_{\alpha } &=0;  \\
\mathcal{Q}_{r} &:I_{2}\left( \mathcal{Q}\right) =0.
\end{array}
\end{equation}
By plugging \eref{eq:b1to3} into \eref{eq:I3-I2-el}, one obtains
\begin{equation}\label{eq:b4}
\left\{
\begin{array}{c@{\ =\ }l}
I_{3}\left( Q\right) &0; \\
\frac{\partial I_{3}\left( Q\right) }{\partial \mathcal{Q}_{r}}&\frac{1}{2} q_{z}\frac{\partial I_{2}\left( \mathcal{Q}\right) }{\partial \mathcal{Q}_{r}},
\end{array}
\right.
\end{equation}
where $\frac{\partial I_{3}\left( Q\right) }{\partial \mathcal{Q}_{r}}$ is the unique possibly non-vanishing component of $\frac{\partial I_{3}\left( Q\right) }{\partial Q_{i}}$.

As above, depending on the vanishing or not of the magnetic charge of Kaluza-Klein vector, the treatment splits as follows:

\begin{enumerate}
\item  In the case
\begin{equation}
q_{z}=0,
\end{equation}
Eq. \eref{eq:b4} yields
\begin{equation}\label{eq:b5}
\left\{
\begin{array}{cl}
I_{3}\left( Q\right) &=0; \\
\frac{\partial I_{3}\left( Q\right) }{\partial Q_{i}}&=0,
\end{array}
\right.
\end{equation}
corresponding to a 1/2-BPS \emphsmall black hole of $\SUSY=8$, $D=5$ supergravity.

\item  On the other hand,for
\begin{equation}
q_{z}\neq 0,
\end{equation}
Eq. \eref{eq:b4} gives
\begin{equation}\label{eq:b6}
\left\{
\begin{array}{cl}
I_{3}\left( Q\right) &=0; \\
\frac{\partial I_{3}\left( Q\right) }{\partial Q_{i}}&\neq 0,
\end{array}
\right.
\end{equation}
corresponding to a 1/4-BPS \emphsmall black hole of $\SUSY=8$, $D=5$ supergravity.
\end{enumerate}
\end{subequations}

Thus, at the level of charge orbits, Eqs. \eref{eq:b1to6} correspond to the following picture:
\begin{equation}\label{eq:b7}\fl\quad
\begin{array}{c}
\orbitfour{2}{small}{1}{6}=\frac{SO(5,5; \fld{R}) }{SO\left( 4,4;\fld{R}\right) \rtimes \fld{R}^{8}} \\
\overset{\displaystyle\swarrow}{\orbittwo{2}{5}=\frac{E_{6\left( 6\right)}(\fld{R})}{SO\left(5,5;\fld{R}\right) \rtimes \fld{R}^{16}}}
\qquad\qquad
\overset{\displaystyle\searrow}{\orbittwo{4}{5}=\frac{E_{6(6)}(\fld{R})}{SO\left( 5,4;\fld{R}\right) \rtimes \fld{R}^{16}}.}
\end{array}
\end{equation}
Indeed, at the level of the semi-simple part of the orbit stabilizers, the following embeddings trivially hold:
\begin{equation}
\begin{array}{c}
SO\left( 4,4;\fld{R}\right) \maxsymmsubsetneq SO\left( 5,4;\fld{R}\right)\maxsymmsubsetneq SO(5,5; \fld{R}) ; \\
SO\left( 4,4;\fld{R}\right) \times SO\left( 1,1;\fld{R}\right) \maxsymmsubsetneq SO(5,5; \fld{R}) .
\end{array}
\end{equation}
At the level of corresponding moduli spaces, \eref{eq:b7} implies that
\begin{equation}
\modulifour{2}{small}{1}{6}=\frac{SO(4,4;\fld{R})}{SO(4,\fld{R})\times SO(4,\fld{R})}\rtimes\fld{R}^{8}
\end{equation}
satisfies (the embedding in the second line being trivial; see also \eref{eq:PPPP-1}) \cite{Cerchiai:2010xv}
\begin{equation}\label{eq:emb-2}\fl
\begin{array}{rl}
\modulifour{2}{small}{1}{6}&\subsetneq \modulitwo{4}{5}=\modulifour{4}{large}{1}{6}\rtimes \fld{R}^{16}\\
&\subsetneq \modulitwo{2}{5}=M_{D=6}\rtimes \fld{R}^{16}\\
&=\displaystyle\frac{SO\left(5,5;\fld{R}\right) }{SO\left( 5, \fld{R}\right) \times SO\left( 5, \fld{R}\right)}\rtimes \fld{R}^{16}.
\end{array}
\end{equation}

Thus, it follows that a \emphsmall 1/2-BPS black hole, as well as a \emphsmall 1/4-BPS black hole, of $\SUSY=8$, $D=5$ supergravity can be uplifted to a \emphsmall 1/2-BPS dyonic black string of $\SUSY=\left( 2,2\right) $, $D=6$ supergravity.

\subsubsection{\texorpdfstring{The 1/4-BPS black hole  orbit under $D=6$ $\to 5$}{The 1/4-BPS black hole  orbit under D=6 -> 5}}\label{sec:p=0,2-1/4}

A representative of the 1/4-BPS \emphsmall charge orbit \eref{eq:PA-13} of a black hole in $\SUSY=\left( 2,2\right) $, $D=6$ supergravity is provided by
\begin{subequations}\label{eq:c1to4}
\begin{equation}\label{eq:c1to2}
\begin{array}{rl}
P^{i}           &=0;  \\
\mathcal{Q}_{r} &=0
\end{array}
\end{equation}
and
\begin{equation}\label{eq:c3}
q_{\alpha }:\left( \gamma ^{r}\right) ^{\alpha \beta }q_{\alpha
}q_{\beta }\neq 0,
\end{equation}
for \emph{at least} some $r=1,...,10$. By plugging \eref{eq:c1to2}, \eref{eq:c3} into \eref{eq:I3-I2-el}, one obtains
\begin{equation}\label{eq:c4}
\left\{
\begin{array}{rl}
I_{3}\left( Q\right) &=0; \\
\frac{\partial I_{3}\left( Q\right) }{\partial \mathcal{Q}_{r}}&=\half\left( \gamma ^{r}\right) ^{\alpha \beta }q_{\alpha }q_{\beta }\neq0,
\end{array}
\right.
\end{equation}
\end{subequations}
where $\frac{\partial I_{3}\left( Q\right) }{\partial\mathcal{Q}_{r}}$ is the unique non-vanishing component of $\frac{\partial I_{3}\left( Q\right) }{\partial Q_{i}}$.

Thus, \emph{independently on the vanishing or not of} $q_{z}$, this case corresponds to a 1/4-BPS \emphsmall black hole of $\SUSY=8$, $D=5$ supergravity.

At the level of charge orbits, Eqs. \eref{eq:c1to4} depict the following situation:
\begin{equation}\label{eq:c5}
\begin{array}{c}
\orbitfour{4}{small}{0,2}{6}=\displaystyle\frac{SO(5,5; \fld{R}) }{SO\left( 3,4;\fld{R}\right) \rtimes \fld{R}^{8}} \\
\downarrow \\
\orbittwo{4}{5}=\displaystyle\frac{E_{6\left( 6\right)}(\fld{R})}{SO\left( 5,4;\fld{R}\right) \rtimes \fld{R}^{16}}.
\end{array}
\end{equation}
Indeed, at the level of the semi-simple part of the orbit stabilizers, the following embedding trivially holds:
\begin{equation}
SO(3,4;\fld{R})\times\left\{\begin{array}{l}SO( 1,1;\fld{R})\\SO(2,\fld{R})\end{array}\right.\maxsymmsubsetneq SO( 5,4;\fld{R}) .
\end{equation}
At the level of corresponding moduli spaces, \eref{eq:c5} implies that
\begin{equation}
\modulifour{4}{small}{0,2}{6}=\frac{SO\left( 3,4;\fld{R}\right) }{SO\left( 3, \fld{R}\right) \times SO\left( 4, \fld{R}\right) }\rtimes\fld{R}^8
\end{equation}
satisfies (see also \eref{eq:PPPP-1}) \cite{Cerchiai:2010xv}
\begin{equation}\label{eq:emb-3}\fl
\begin{array}{rl}
\modulifour{4}{small}{0,2}{6}&\subsetneq\modulitwo{4}{5}=\modulifour{4}{large}{1}{6}\rtimes \fld{R}^{16}\\
&=\displaystyle\frac{SO\left( 5,4; \fld{R}\right) }{SO\left( 5, \fld{R}\right) \times SO\left( 4, \fld{R}\right) }\rtimes \fld{R}^{16}.
\end{array}
\end{equation}

Thus, it follows that a \emphsmall 1/4-BPS black hole of $\SUSY=8$, $D=5$ supergravity can be uplifted to a \emphsmall 1/4-BPS black hole of $\SUSY=\left( 2,2\right) $, $D=6$ supergravity.

\subsubsection{\texorpdfstring{The 1/2-BPS black hole  orbit under $D=6$ $\to 5$}{The 1/2-BPS black hole  orbit under D=6 -> 5}}\label{sec:p=0,2-1/2}

A representative of the 1/2-BPS \emphsmall charge orbit \eref{eq:PA-14} of a black hole in $\SUSY=\left( 2,2\right) $, $D=6$
supergravity is provided by
\begin{subequations}\label{eq:d1to4}
\begin{equation}\label{eq:d1to2}
\begin{array}{rl}
P^{i}           &=0;   \\
\mathcal{Q}_{r} &=0
\end{array}
\end{equation}
and
\begin{equation}\label{eq:d3}
q_{\alpha }:\left( \gamma ^{r}\right) ^{\alpha \beta }q_{\alpha}q_{\beta }=0,~\forall r.
\end{equation}
By plugging \eref{eq:d1to2}, \eref{eq:d3} into \eref{eq:I3-I2-el}, one obtains
\begin{equation}\label{eq:d4}
\left\{
\begin{array}{rl}
I_{3}\left( Q\right) &=0; \\
\frac{\partial I_{3}\left( Q\right) }{\partial \mathcal{Q}_{r}}&=\frac{1}{2}\left( \gamma ^{r}\right) ^{\alpha \beta }q_{\alpha }q_{\beta}=0\Leftrightarrow \frac{\partial I_{3}\left( Q\right) }{\partial Q_{i}}=0.
\end{array}
\right.
\end{equation}
\end{subequations}

Thus, \emph{independently on the vanishing or not of} $q_{z}$, this case corresponds to a 1/2-BPS \emphsmall black hole of $\SUSY=8$, $D=5$ supergravity.

At the level of charge orbits, Eqs. \eref{eq:d1to4} depict the following situation:
\begin{equation}\label{eq:d5}
\begin{array}{c}
\orbitfour{2}{small}{0,2}{6}=\displaystyle\frac{SO(5,5; \fld{R}) }{SL\left( 5,\fld{R}\right) \rtimes \fld{R}^{10}}  \\
\downarrow  \\
\orbittwo{2}{5}=\displaystyle\frac{E_{6\left( 6\right)}(\fld{R})}{SO(5,5; \fld{R}) \rtimes \fld{R}^{16}}.
\end{array}
\end{equation}
Indeed, at the level of semi-simple part of the stabilizers of orbits, the following embedding trivially holds:
\begin{equation}
SL\left( 5,\fld{R}\right) \times SO\left( 1,1; \fld{R}\right) \maxsymmsubsetneq SO(5,5; \fld{R}) .
\end{equation}
At the level of corresponding moduli spaces, \eref{eq:d5} implies that
\begin{equation}
\modulifour{2}{small}{0,2}{6}=M_{D=7}\rtimes\fld{R}^{10}=\frac{SL\left( 5,\fld{R}\right) }{SO\left( 5, \fld{R}\right) }\rtimes\fld{R}^{10}
\end{equation}
satisfies \cite{Cerchiai:2010xv}
\begin{equation}\label{eq:emb-4}
\modulifour{2}{small}{0,2}{6}\subsetneq ~\modulitwo{2}{5}=M_{D=6}\rtimes\fld{R}^{16},
\end{equation}
which is a trivial consequence of the embedding between the scalar manifolds of maximal supergravity in $D=6$ and $D=7$.

Thus, it follows that a \emphsmall 1/2-BPS black hole of $\SUSY=8$, $D=5$ supergravity can be uplifted to a \emphsmall 1/2-BPS black hole of $\SUSY=\left( 2,2\right) $, $D=6$ supergravity.

Summarizing the embeddings of moduli spaces \eref{eq:emb-1}, \eref{eq:emb-2}, \eref{eq:emb-3} and \eref{eq:emb-4}, related to the various $D=6\to 5$ dimensional reductions considered in \autoref{sec:p=1-1/4}-\autoref{sec:p=0,2-1/2}, the following result is achieved achieved \cite{Cerchiai:2010xv}:
\begin{equation}\label{eq:emb-d=5-d=6-tot}\fl
\begin{array}{rl}
\modulifour{4}{small}{0,2}{6}&\maxsymmsubsetneq \modulifour{2}{small}{1}{6}\\
&\subsetneq \modulitwo{4}{5} =\modulifour{4}{large}{1}{6}\rtimes\fld{R}^{16}\\
&\left\{\begin{array}{l}
\subsetneq\modulitwo{8}{5}; \\
\maxsymmsubsetneq\modulitwo{2}{5}\supsetneq\modulifour{2}{small}{0,2}{6}.
\end{array}\right.
\end{array}
\end{equation}

\subsection{\texorpdfstring{U-duality orbits of $E_{6(6)}(\rng{Z})$}{U-duality orbits of E\_6(6)(Z)}}\label{sec:D5disc}

For quantized charges the continuous  U-duality is broken to an infinite discrete subgroup, which for $D=5$ is given by $E_{6(6)}(\rng{Z})\subset E_{6(6)}(\fld{R})$ \cite{Hull:1994ys}. The integral Jordan algebra $\mathfrak{J}^{\fld{O}_{\rng{Z}}^s}_{3}$ of integral split-octonionic $3\times 3$ Hermitian matrices provides a natural model for $E_{6(6)}(\rng{Z})$, which may used to analyse the discrete U-duality orbits. See \cite{Gross:1996,Elkies:1996,Krutelevich:2002} and \ref{sec:5DJ} for further details. The quantized black hole  charge vector is given by,
\begin{equation}\fl\qquad
Q=\left(\begin{array}{ccc}q_1&Q_s&\overline{Q_c}\\\overline{Q_s}&q_2&Q_v\\Q_c&\overline{Q}_v&q_3\end{array}\right), \ \  \mathrm{where} \ \  q_1,q_2, q_3 \in \rng{Z} \ \  \mathrm{and} \ \  Q_{v,s,c}\in\fld{O}^{s}_{\rng{Z}}.
\end{equation}
The discrete group $E_{6(6)}(\rng{Z})$ is defined by the set of norm-preserving invertible $\rng{Z}$-linear transformations,
\begin{equation}
\{\sigma:\mathfrak{J}^{\fld{O}_{\rng{Z}}^s}_{3}\to\mathfrak{J}^{\fld{O}_{\rng{Z}}^s}_{3}|N_3(\sigma(Q))=N_3(Q)\}.
\end{equation}
It is with this framework that we shall study the discrete U-duality orbits.

As was the case in $D=6$, the charge conditions defining the orbits in the continuous theory are manifestly invariant under the discrete subgroup $E_{6(6)}(\rng{Z})$ and, hence, those states unrelated by U-duality in the classical theory remain unrelated in the quantum theory.   There are three disjoint \emph{classes} of orbits, one 1/2-BPS, one 1/4-BPS and one 1/8-BPS, corresponding to the three continuous orbits.   However, each of these classes is broken up into a countably infinite set of discrete orbits \cite{Borsten:2009zy}. To classify these orbits Krutelevich used $E_{6(6)}(\rng{Z})$ to bring an arbitrary charge vector into a \emph{diagonal reduced} canonical form, which is uniquely defined by the following set of three discrete invariants,
\begin{equation}\label{eq:5dinvs}
\begin{array}{r@{\ :=\ }l}
c_1(Q)&\gcd(Q),\\
c_2(Q)&\gcd(Q^\sharp),\\
c_3(Q)&N_3(Q).
\end{array}
\end{equation}
See \cite{Krutelevich:2002} and \ref{sec:33Jz} for details. Note, $c_2$ is only well defined for black holes preserving less than 1/2 the supersymmetries and hence only enters the discrete orbit classification for 1/4-BPS and 1/8-BPS states. Further, the invariants $c_i$ appear in the 1/8-BPS degeneracy formula, for primitive states, of \cite{Maldacena:1999bp}.

\paragraph*{$D=5$ diagonal reduced canonical form (Krutelevich 2002 \cite{Krutelevich:2002})} Every element $Q\in\mathfrak{J}^{\fld{O}^{s}_{\rng{Z}}}_{3}$ is $E_{6(6)}(\rng{Z})$ equivalent to a diagonally reduced canonical form,
\begin{equation}\label{eq:5dcan}
Q_{\mathrm{can}}=k\left(\begin{array}{ccc}1&0&0\\0&l&0\\0&0&lm\end{array}\right),
\ \  \mathrm{where} \ \  k>0, l\geq 0.
\end{equation}
The canonical form is uniquely determined by \eref{eq:5dinvs} since
\begin{equation}
\begin{array}{r@{\ =\ }l}
c_1(Q_{\mathrm{can}})&k,\\
c_2(Q_{\mathrm{can}})&k^2l,\\
c_3(Q_{\mathrm{can}})&k^3l^2m.
\end{array}
\end{equation}
so that for arbitrary $Q$ one obtains $k=c_1(Q)$, $l=k^{-2}c_2(Q)$ and $m=k^{-3}l^{-2}c_3(Q)$.

\paragraph{$D=5$ Black hole orbit classification}
\begin{enumerate}
\item The complete set of distinct 1/2-BPS charge vector orbits is given by,
\begin{equation}
\big\{\left(\begin{array}{ccc}k&0&0\\0&0&0\\0&0&0\end{array}\right), \ \
\mathrm{where} \ \  k>0 \big\}.
\end{equation}

\item The complete set of distinct 1/4-BPS charge vector orbits is given by,
\begin{equation}
\big\{\left(\begin{array}{ccc}k&0&0\\0&kl&0\\0&0&0\end{array}\right), \ \
\mathrm{where} \ \  k,l>0 \big\}.
\end{equation}
\item The complete set of distinct 1/8-BPS charge vector orbits is given by,
\begin{equation}
\big\{\left(\begin{array}{ccc}k&0&0\\0&kl&0\\0&0&klm\end{array}\right), \ \
\mathrm{where} \ \  k,l,|m|>0 \big\}.
\end{equation}
\end{enumerate}

\section{\texorpdfstring{Black holes in $D=4$}{Black holes in D=4}}\label{sec:D4}

\subsection{\texorpdfstring{U-duality orbits of $E_{7(7)}(\fld{R})$}{U-duality orbits of E\_7(7)(R)}}\label{sec:D4cont}

In the classical supergravity limit the 28+28 electric/magnetic black hole charges $x_I$ ($I=1,\ldots ,56$ throughout) transform as the fundamental \rep{56} of the continuous U-duality group $E_{7(7)}(\fld{R})$. Under $SO(1,1; \fld{R})\times E_{6(6)}(\fld{R})$ the \rep{56} breaks as
\begin{equation}\label{eq:56-branching}
\mathbf{56}\to
\mathbf{1}_{3}+\mathbf{1}_{-3}+\mathbf{27}_{1}+\mathbf{27'}_{-1}
\end{equation}
where the singlets may be identified as the graviphoton charge and its electromagnetic dual descending from $D=5$. In this basis the charges $x_I$ may be conveniently represented as an element $x$ of the Freudenthal triple system
$\mathfrak{M}(\mathfrak{J}^{\fld{O}^s}_{3})$,
\begin{equation}
x=\left(\begin{array}{cc}-q_0&P\\Q&p^0\end{array}\right), \ \  \mathrm{where}\ \  q_0, p^0 \in \fld{R} \ \  \mathrm{and} \ \  Q,P\in\mathfrak{J}^{\fld{O}^s}_{3}.
\end{equation}
Here, $p^0, q_0$ are the graviphotons and $P,Q$ are the magnetic/electric $\mathbf{27}'$ and \rep{27} respectively. The quartic norm is given by,
\begin{equation}\label{eq:I4-branching}\fl\qquad
\Delta(x)=-[p^0q_0+\Tr(P,Q)]^2+4[q_0
N(P)-p^0 N(Q)+\Tr(P^\sharp, Q^\sharp)].
\end{equation}
See \ref{sec:5DJ} and \ref{sec:fts} for the necessary definitions.

The set of invertible linear transformations leaving the quartic norm and the antisymmetric bilinear form \eref{eq:bilinearform} invariant is nothing but the $D=4$ U-duality group $E_{7(7)}(\fld{R})$. Moreover,
\begin{equation}
\Delta(x)=I_4(x),
\end{equation}
where,
\begin{equation}
I_4(Q)=d^{IJKL}x_I x_J x_K x_L
\end{equation}
and $d^{IJKL}$ is the $E_{7(7)}(\fld{R})$ invariant tensor. The black hole entropy is simply given by the quartic norm \cite{Kallosh:1996uy},
\begin{equation}
S_{D=4, \mathrm{BH}}=\pi\sqrt{|\Delta(x)|}=\pi\sqrt{|I_4(x)|}.
\end{equation}

In this case there are five U-duality orbits, three 1/2-BPS, 1/4-BPS, 1/8-BPS \emphsmall orbits and two \emphlarge orbits, one 1/8-BPS and one non-BPS depending on the  sign of the unique quartic $E_{7(7)}(\rng{Z})$ invariant \cite{Ferrara:1997uz}. These orbits may distinguished by the FTS rank of $x$,
\begin{equation}\fl\qquad
\begin{tabular}{@{Rank }c@{\quad}M{c}@{\quad}c@{-BPS,}}
1 & x\not=0, 3T(x,x,y)+x\{x,y\}=0 \, \forall y                   & 1/2\\[3pt]
2 & \exists y\,\mathrm{s.t.}\,3T(x,x,y)+x\{x,y\}\not=0, T(x,x,x)=0 & 1/4\\[3pt]
3 & T(x,x,x)\not=0,\Delta(x)=0                                   & 1/8\\[3pt]
4 & \Delta(x)>0                                                  & 1/8\\[3pt]
4 & \Delta(x)<0                                                  & non
\end{tabular}
\end{equation}
where $T(x,x,x)$ is the triple product \eref{eq:Tofx} and $\{x,y\}$ is the antisymmetric bilinear form \eref{eq:bilinearform}. See \ref{sec:fts} for details.

Note, $T(x,x,x)$ transforms as a \rep{56} under $E_{7(7)}$. Similarly
\begin{equation}
\partial_I I_4(Q)=\frac{\partial I_4(Q)}{\partial x_I}=2d^{IJKL}x_Jx_Kx_L
\end{equation}
manifestly transforms as a \rep{56} under $E_{7(7)}$ so that $2T(x,x,x)=\partial_I I_4(Q)$. Moreover, $3T(x,x,y)+x\{x,y\}$ vanishes for all $y$ if and only if the $\rep{133}$ in $\rep{56}\times_s\rep{56}$ vanishes. Hence, the FTS rank conditions are equivalent to those originally used in \cite{Ferrara:1997ci},
\begin{equation}
\begin{tabular}{@{Rank }c@{\quad}M{c}@{\quad}c@{-BPS}l}
1 & x\not=0, \partial^{2}_{IJ}I_4(x)|_{133}=0                 & 1/2&,\\[3pt]
2 & \partial^{2}_{IJ}I_4(x)|_{133}\not=0, \partial_I I_4(x)=0 & 1/4&,\\[3pt]
3 & \partial_I I_4(x)\not=0, I_4(x)=0                         & 1/8&,\\[3pt]
4 & I_4(x)>0                                                  & 1/8&,\\[3pt]
4 & I_4(x)<0                                                  & non&.
\end{tabular}
\end{equation}
The orbits with their rank conditions, dimensions and representative states are summarized in \autoref{tab:Frank}.
\begin{table}
\caption[$D=4$ black hole orbits, their corresponding rank conditions, dimensions and SUSY.]{$D=4$ black hole orbits, their corresponding rank conditions, dimensions and SUSY.\label{tab:Frank}}
\begingroup
\begin{tabular*}{\textwidth}{@{\extracolsep{\fill}}F{c}*{4}{T{c}}F{c}F{c}}
\toprule
\multirow{2}{*}{Rank} & \multicolumn{2}{c}{\small{Rank/orbit conditions}} &\multirow{2}{*}{Rep state} & \multirow{2}{*}{Orbit} & \multirow{2}{*}{dim} & \multirow{2}{*}{SUSY}  \\
\cmidrule(r){2-3}
& \mathrm{non\mbox{-}vanishing\ }(\exists\ y\ \mathrm{s.t.})& \mathrm{vanishing\ }(\forall\ y) \\
\midrule
1 & x                  & 3T(x,x,y)+x\{x,y\} & \left(\begin{array}{cc}1&(0,0,0)\\0&0\end{array}\right)& \frac{E_{7(7)}(\fld{R})}{E_{6(6)}(\fld{R})\ltimes\fld{R}^{27}} & 28 & 1/2  \\
2 & 3T(x,x,y)+x\{x,y\} & T(x,x,x)           & \left(\begin{array}{cc}1&(1,0,0)\\0&0\end{array}\right)& \frac{E_{7(7)}(\fld{R})}{O(6,5;\fld{R})\ltimes\fld{R}^{32}\times\fld{R}} & 45 & 1/4  \\
3 & T(x,x,x)           & \Delta(x)          & \left(\begin{array}{cc}0&(1,1,1)\\0&0\end{array}\right)& \frac{E_{7(7)}(\fld{R})}{F_{4(4)}(\fld{R})\ltimes\fld{R}^{26}} & 55 & 1/8 \\
4 & \Delta(x)>0        & -                  & \left(\begin{array}{cc}1&(1,1,k)\\0&0\end{array}\right)& \frac{E_{7(7)}(\fld{R})}{E_{6(2)}(\fld{R})} & 55 & 1/8 \\
4 & \Delta(x)<0        & -                  & \left(\begin{array}{cc}1&(0,0,0)\\0&k\end{array}\right)& \frac{E_{7(7)}(\fld{R})}{E_{6(6)}(\fld{R})} & 55 & 0\\
\bottomrule
\end{tabular*}
\endgroup
\end{table}

\subsection{\texorpdfstring{$D=4 \longleftrightarrow D=5$ relations for charge orbits and moduli spaces}{D=4 <-> D=5 relations for charge orbits and moduli spaces}}\label{sec:4/5-Rels}

The decomposition \eref{eq:56-branching} of the \rep{56} irrep of $D=4$ U-duality $E_{7\left( 7\right) }(\fld{R})$ with respect to $D=5$ U-duality $E_{6(6)}(\fld{R})\left( \times SO\left( 1,1;\fld{R}\right) \right)$ \cite{Ferrara:1997ci,Ferrara:1997uz,Lu:1997bg,Ceresole:2007rq,Bianchi:2009mj} corresponds to the following branching of the $D=4$ charge vector $x_I$:
\begin{equation}
x_I\to \left( q_{0},Q_{i},P^{i},p^{0}\right) .
\end{equation}
Thus, the unique invariant $I_{4}$ of the \rep{56} can be written in a manifestly $E_{6(6)}(\fld{R})$-invariant way as follows \cite{Ferrara:1997uz} (cf. Eq. \eref{eq:I4-branching}):
\begin{equation}\label{eq:I4}\fl\qquad
I_{4}=-\left( p^{0}q_{0}+P^{i}Q_{i}\right) ^{2}+4\left[q_{0}I_{3}\left(P\right) -p^{0}I_{3}\left( Q\right) +\frac{\partial I_{3}\left( Q\right) }{\partial Q_{i}}\frac{\partial I_{3}\left( P\right) }{\partial P^{i}}\right],
\end{equation}
where $q_{0}$ and $p^{0}$ respectively are the electric and magnetic charge of the $d=5\to 4$ Kaluza-Klein vector, and the unique cubic electric (magnetic) invariant of the \rep{27} (\rep{27'}) irrep of $E_{6(6)}(\fld{R})$ is defined as follows \cite{Ferrara:1997uz,Ferrara:1997ci}:
\begin{subequations}
\begin{eqnarray}
I_{3}\left( q\right)  &:=\frac{1}{3!}d^{ijk}Q_{i}Q_{j}Q_{k};\label{eq:I3-el}\\
I_{3}\left( p\right)  &:=\frac{1}{3!}d_{ijk}P^{i}P^{j}P^{k}.\label{eq:I3-magn}
\end{eqnarray}
\end{subequations}

By recalling the \emph{continuous} U-invariant definitions of the \emphlarge and \emphsmall charge orbits of $\SUSY=8$ supergravity in $D=4$ and $D=5$ \cite{Ferrara:1997uz,Lu:1997bg,D'Auria:1999fa}, we are now going to elucidate the various relations among them (and the consequences for the related moduli spaces), by performing a $D=5$ $\to 4$ spacelike dimensional reduction, with vanishing or non-vanishing Kaluza-Klein charges. See also \cite{Ceresole:2009jc,Ceresole:2009id,Bianchi:2009mj}.

\subsubsection{\texorpdfstring{The 1/8-BPS orbit under $D=5$ $\to 4$}{The 1/8-BPS orbit under D=5 -> 4}}\label{sec:1/8}

A representative of 1/8-BPS charge orbit of $\SUSY=8$, $D=5$ supergravity \cite{Ferrara:1997uz,Lu:1997bg}
\begin{equation}
\orbittwo{8}{5}=\frac{E_{6\left( 6\right)}(\fld{R})}{F_{4\left( 4\right) }(\fld{R})}
\end{equation}
is provided by a black hole with (\emphlarge $3$-charges solution)
\begin{subequations}\label{eq:1to5}
\begin{equation}\label{eq:1to2}
\begin{array}{rl}
P^{i} &=0;   \\
Q_{i} &:I_{3}\left( Q\right) \neq 0,
\end{array}
\end{equation}
with non-vanishing Bekenstein-Hawking entropy\footnote{Notice that in $D=5$ $\opname{sgn}\left( I_{3}\right) $ is not relevant since it flips under CPT transformations. The relative signs of the three charges in the diagonally reduced form, and hence $\opname{sgn}(I_3)$, play no role - all large black holes are 1/8-BPS \cite{Ferrara:1997uz,Ferrara:1997ci}.} given by Eq. \eref{eq:D=5-entropy}.

By setting
\begin{equation}\label{eq:q0=0}
q_{0}=0
\end{equation}
and plugging \eref{eq:1to2} into \eref{eq:I4}, one obtains
\begin{equation}\label{eq:3}
I_{4}=-4p^{0}I_{3}\left( Q\right).
\end{equation}

Depending on the vanishing or not of the magnetic charge of Kaluza-Klein vector, the treatment splits as follows:

\begin{enumerate}
\item  In the case
\begin{equation}
p^{0}=0,
\end{equation}
Eq. \eref{eq:3} yields
\begin{equation}\label{eq:4}
\left\{
\begin{array}{rl}
I_{4}&=0; \\
\frac{\partial I_{4}}{\partial p^{0}}&=4I_{3}\left( Q\right) \neq 0,
\end{array}
\right.
\end{equation}
corresponding to a 1/8-BPS \emphsmall black hole of $\SUSY=8$, $D=4$ supergravity.

\item  On the other hand, if
\begin{equation}
p^{0}\neq 0,
\end{equation}
then Eq. \eref{eq:3} yields\footnote{Notice that $\opname{sgn}\left( I_{4}\right)$ is a U-duality invariant quantity which distinguishes between 1/8-BPS and non-BPS large black holes \cite{Ferrara:1997uz,Ferrara:1997ci}.}
\begin{equation}\label{eq:5}
I_{4}\left\{\begin{array}{ll}
>0&\mathrm{if\ }\opname{sgn}\left( p^{0}I_{3}\left( Q\right) \right) =-1; \\
<0&\mathrm{if\ }\opname{sgn}\left( p^{0}I_{3}\left( P\right) \right) =1,
\end{array}\right.
\end{equation}
thus corresponding to a 1/8-BPS ($I_{4}>0$) or to a non-BPS $Z_{AB}\neq 0$ ($I_{4}<0$) \emphlarge black hole of $\SUSY=8$, $D=4$ supergravity.
\end{enumerate}
\end{subequations}

Therefore, at the level of charge orbits, Eqs. \eref{eq:1to5} correspond to the following picture:
\begin{equation}\label{eq:6}\fl\qquad
\begin{array}{c}
\mathcal{O}_{\mathrm{1/8-BPS, small, }D=4}=\frac{E_{7\left( 7\right) }(\fld{R})}{F_{4\left( 4\right) }(\fld{R})\rtimes \fld{R}^{26}} \\
\uparrow \\
\orbittwo{8}{5}=\frac{E_{6\left( 6\right)}(\fld{R})}{F_{4\left(4\right) }(\fld{R})} \\
\overset{\displaystyle\swarrow}{\mathcal{O}_{\mathrm{1/8-BPS, large, }D=4}=\frac{E_{7\left( 7\right) }(\fld{R})}{E_{6\left( 2\right) }(\fld{R})}}
\qquad\qquad
\overset{\displaystyle\searrow}{\mathcal{O}_{\mathrm{non\mbox{-}BPS, }D=4}=\frac{E_{7\left( 7\right) }(\fld{R})}{E_{6(6)}(\fld{R})}.}
\end{array}
\end{equation}
Indeed, at the level of the semi-simple part of the orbit stabilizers, it holds that
\begin{equation}
\begin{array}{rl}
F_{4\left( 4\right) }(\fld{R}) &\maxsymmsubsetneq E_{6\left( 2\right) }(\fld{R}); \\
F_{4\left( 4\right) }(\fld{R}) &\maxsymmsubsetneq E_{6\left(6\right) }(\fld{R}).
\end{array}
\end{equation}
At the level of corresponding moduli spaces, \eref{eq:6} implies that
\begin{equation}\label{eq:7}
\modulitwo{8}{5}=\frac{F_{4\left( 4\right) }(\fld{R})}{USp\left( 6,\fld{R}\right) \times USp\left(2,\fld{R}\right) }
\end{equation}
satisfies (the first embedding being trivial)
\begin{equation}\label{eq:8}\fl\qquad
\modulitwo{8}{5}\subsetneq \left\{\begin{array}{l}
\displaystyle\mathcal{M}_{\mathrm{non\mbox{-}BPS, }D=4}=M_{D=5}=\frac{E_{6(6)}(\fld{R})}{USp\left( 8,\fld{R}\right) }; \\
\displaystyle\mathcal{M}_{\mathrm{1/8-BPS, large, }D=4}=\frac{E_{6\left(2\right) }(\fld{R})}{SU\left( 6,\fld{R}\right)\times SU\left( 2,\fld{R}\right) },
\end{array}\right.
\end{equation}
yielding the following relation:
\begin{equation}
\modulitwo{8}{5}\subset \left[ M_{D=5}\cap \mathcal{M}_{\mathrm{1/8-BPS, large, }D=4}\right] .
\end{equation}

Thus, it follows that a \emphsmall 1/8-BPS black hole, as well as a \emphlarge 1/8-BPS and a \emphlarge non-BPS black hole, of $\SUSY=8$, $D=4$ supergravity can be uplifted to a \emphlarge 1/8-BPS black hole of $\SUSY=8$, $D=5$ supergravity.

\subsubsection{\texorpdfstring{The 1/4-BPS orbit under $D=5$ $\to 4$}{The 1/4-BPS orbit under D=5 -> 4}}\label{sec:1/4}

A representative of 1/4-BPS charge orbit of $\SUSY=8$, $D=5$ supergravity \cite{Ferrara:1997uz,Lu:1997bg}
\begin{equation}
\orbittwo{4}{5}=\frac{E_{6\left( 6\right)}(\fld{R})}{SO\left( 5,4;\fld{R}\right) \rtimes \fld{R}^{16}}
\end{equation}
is provided by a black hole with (\emphsmall 2-charges solution)
\begin{subequations}\label{eq:1-1to5}
\begin{equation}\label{eq:1-1to2}
\begin{array}{rl}
P^{i} &=0;   \\
Q_{i} &:
\left\{\begin{array}{rl}
I_{3}\left( Q\right) &=0; \\
\displaystyle\frac{\partial I_{3}\left( Q\right) }{\partial Q_{i}}&=\half d^{ijk}Q_{j}Q_{k}\neq 0,
\end{array}\right.
\end{array}
\end{equation}
thus with vanishing Bekenstein-Hawking entropy given by Eq. \eref{eq:D=5-entropy}.

By setting \eref{eq:q0=0} and plugging \eref{eq:1-1to2} into \eref{eq:I4}, one obtains
\begin{equation}\label{eq:1-3}
\left\{
\begin{array}{rl}
I_{4}&=0; \\
\frac{\partial I_{4}}{\partial Q_{i}}&=4p^{0}\frac{\partial I_{3}\left( Q\right) }{\partial Q_{i}},
\end{array}
\right.
\end{equation}
where $\frac{\partial I_{4}}{\partial Q_{i}}$ is the unique possibly non-vanishing component of $\frac{\partial I_{4}}{\partial x_I}$.

Depending on the vanishing or not of the magnetic charge of Kaluza-Klein vector, the treatment splits as follows:

\begin{enumerate}
\item  In the case
\begin{equation}
p^{0}=0,
\end{equation}
Eq. \eref{eq:1-3} yields
\begin{equation}\label{eq:1-4}
\left\{
\begin{array}{rl}
I_{4}&=0; \\
\frac{\partial I_{4}}{\partial x}&=0, \\
\frac{\partial ^{2}I_{4}}{\partial Q_{i}\partial p^{0}}&=4\frac{\partial I_{3}\left( Q\right) }{\partial Q_{i}}\neq 0,
\end{array}
\right.
\end{equation}
where $\frac{\partial ^{2}I_{4}}{\partial Q_{i}\partial p^{0}}$ is the unique non-vanishing component of $\left. \frac{\partial
^{2}I_{4}}{\partial x_I\partial x_J}\right| _{\rep{133}}$. Therefore, this case corresponds to a 1/4-BPS \emphsmall black hole of $\SUSY=8$, $D=4$ supergravity.

\item  On the other hand, for
\begin{equation}
p^{0}\neq 0
\end{equation}
Eq. \eref{eq:1-3} yields
\begin{equation}\label{eq:1-5}
\left\{
\begin{array}{rl}
I_{4}&=0; \\
\frac{\partial I_{4}}{\partial x_I}&\neq 0,
\end{array}
\right.
\end{equation}
corresponding to a \emphsmall 1/8-BPS black hole of $\SUSY=8$, $D=4$ supergravity.
\end{enumerate}
\end{subequations}

Therefore, at the level of charge orbits, Eqs. \eref{eq:1-1to5} correspond to the following picture:
\begin{equation}\label{eq:1-6}\fl
\begin{array}{c}
\orbittwo{4}{5}=\frac{E_{6\left( 6\right)}(\fld{R})}{SO\left(5,4;\fld{R}\right) \rtimes \fld{R}^{16}} \\
\overset{\displaystyle\swarrow}{\orbittwo{4}{4}=\frac{E_{7\left( 7\right)}(\fld{R})}{SO\left(6,5;\fld{R}\right) \rtimes \fld{R}^{32}\times\fld{R}}}
\qquad\qquad
\overset{\displaystyle\searrow}{\mathcal{O}_{\mathrm{1/8-BPS, small, }D=4}=\frac{E_{7\left( 7\right) }(\fld{R})}{F_{4\left( 4\right)(\fld{R})}\rtimes \fld{R}^{26}}.}
\end{array}
\end{equation}
Indeed, at the level of the semi-simple part of the orbit stabilizers, it holds that
\begin{equation}
\begin{array}{c}
SO\left( 5,4;\fld{R}\right)  \maxsymmsubsetneq F_{4\left( 4\right)}(\fld{R});\label{eq:PP-1} \\
SO\left( 5,4;\fld{R}\right)\times\left\{\begin{array}{l}SO( 2,\fld{R})  \\SO( 1,1;\fld{R})\end{array}\right. \maxsymmsubsetneq SO( 6,5;\fld{R}).
\end{array}
\end{equation}
At the level of corresponding moduli spaces, \eref{eq:1-6} implies that \cite{Cerchiai:2010xv}
\begin{equation}\label{eq:PPPP-1}
\modulitwo{4}{5}=\frac{SO\left( 5,4;\fld{R}\right)}{SO\left( 5,\fld{R}\right) \times SO\left( 4,\fld{R}\right)}\rtimes\fld{R}^{16}
\end{equation}
satisfies
\begin{equation}\label{eq:1/4-embs}\fl
\modulitwo{4}{5}\subsetneq
\left\{\begin{array}{l}
\modulitwo{4}{4}=\displaystyle\frac{SO\left( 6,5;\fld{R}\right) }{SO\left( 6,\fld{R}\right) \times SO\left( 5,\fld{R}\right) }\rtimes(\fld{R}^{32}\times\fld{R}); \\
\begin{array}{rl}
\mathcal{M}_{\mathrm{1/8-BPS, small, }D=4}&=\modulitwo{8}{5}\rtimes\fld{R}^{26}\\
&=\displaystyle\frac{F_{4\left( 4\right) }(\fld{R})}{USp\left( 6,\fld{R}\right) \times USp\left( 2,\fld{R}\right) }\rtimes\fld{R}^{26},
\end{array}
\end{array}\right.
\end{equation}
yielding the following relation:
\begin{equation}
\modulitwo{4}{5}\subset \left[ \modulitwo{4}{4}\cap \mathcal{M}_{\mathrm{1/8-BPS, small, }D=4}\right]
.\label{eq:PPP-1}
\end{equation}

Thus, it follows that a \emphsmall 1/4-BPS black hole, as well as a \emphsmall 1/8-BPS black hole, of $\SUSY=8$, $D=4$ supergravity can be uplifted to a \emphsmall 1/4-BPS black hole of $\SUSY=8$, $D=5$ supergravity.

\subsubsection{\texorpdfstring{The 1/2-BPS orbit under $D=5$ $\to 4$}{The 1/2-BPS orbit under D=5 -> 4}}\label{sec:1/2}

A representative of 1/2-BPS charge orbit of $\SUSY=8$, $D=5$ supergravity \cite{Ferrara:1997uz,Lu:1997bg}
\begin{equation}
\orbittwo{2}{5}=\frac{E_{6\left( 6\right)
}(\fld{R})}{SO(5,5; \fld{R}) \rtimes \mathbb{R}^{16}}
\end{equation}
is provided by a black hole with (\emphsmall $1$-charge solution)
\begin{subequations}\label{eq:2-1to5}
\begin{equation}\label{eq:2-1to2}
\begin{array}{rl}
P^{i} &=0;   \\
Q_{i} &:
\left\{\begin{array}{ll}
I_{3}\left( Q\right) &=0; \\
\displaystyle\frac{\partial I_{3}\left( Q\right) }{\partial Q_{i}}&=\half d^{ijk}Q_{j}Q_{k}=0,
\end{array}\right.
\end{array}
\end{equation}
with
\begin{equation}
\frac{\partial ^{2}I_{3}\left( Q\right) }{\partial Q_{i}\partial Q_{j}}=d^{ijk}Q_{k}\neq 0
\end{equation}
for \emph{at least} some $i\in \left\{ 1,...,27\right\} $, due to the non-triviality of the background under consideration. As given by Eq. \eref{eq:D=5-entropy}, the resulting Bekenstein-Hawking entropy vanishes.

By plugging \eref{eq:2-1to2} into \eref{eq:I4}, one obtains
\begin{equation}\label{eq:2-3}
\left\{
\begin{array}{rl}
I_{4}&=0; \\
\frac{\partial I_{4}}{\partial Q_{i}}&=0; \\
\frac{\partial ^{2}I_{4}}{\partial p^{0}\partial Q_{i}}&=4p^{0}\frac{\partial ^{2}I_{3}\left( Q\right) }{\partial Q_{i}\partial Q_{j}},
\end{array}
\right.
\end{equation}
where $4p^{0}\frac{\partial ^{2}I_{3}\left( Q\right) }{\partial Q_{i}\partial Q_{j}}$ is the unique possibly non-vanishing component of $\left. \frac{\partial ^{2}I_{4}}{ \partial x_I\partial x_J}\right| _{\rep{133}}$ .

As above, depending on the vanishing or not of the magnetic charge of Kaluza-Klein vector, the treatment splits as follows:

\begin{enumerate}
\item  In the case
\begin{equation}
p^{0}=0,
\end{equation}
Eq. \eref{eq:2-3} yields
\begin{equation}\label{eq:2-4}
\left\{
\begin{array}{rl}
I_{4}&=0; \\
\frac{\partial I_{4}}{\partial x_I}&=0, \\
\left. \frac{\partial ^{2}I_{4}}{ \partial x_I\partial x_J}\right| _{\rep{133}}&=0.
\end{array}
\right.
\end{equation}
Therefore, this case corresponds to a 1/2-BPS \emphsmall black hole of $\SUSY=8$, $D=4$ supergravity.

\item  On the other hand, for
\begin{equation}
p^{0}\neq 0,
\end{equation}
Eq. \eref{eq:2-3} gives
\begin{equation}\label{eq:2-5}
\left\{
\begin{array}{rl}
I_{4}&=0; \\
\frac{\partial I_{4}}{\partial x_I}&=0, \\
\left. \frac{\partial ^{2}I_{4}}{ \partial x_I\partial x_J}\right| _{\rep{133}}&\neq 0,
\end{array}
\right.
\end{equation}
corresponding to a \emphsmall 1/4-BPS black hole of $\SUSY=8$, $D=4$ supergravity.
\end{enumerate}
\end{subequations}

Thus, at the level of charge orbits, Eqs. \eref{eq:2-1to5} correspond to the following picture:
\begin{equation}\label{eq:2-6}\fl
\begin{array}{c}
\orbittwo{2}{5}=\frac{E_{6\left( 6\right)}(\fld{R})}{SO\left(5,5;\fld{R}\right) \rtimes \fld{R}^{16}} \\
\overset{\displaystyle\swarrow}{\orbittwo{2}{4}=\frac{E_{7\left( 7\right)}(\fld{R})}{E_{6\left(6\right) }(\fld{R})\rtimes \fld{R}^{27}}}
\qquad\qquad
\overset{\displaystyle\searrow}{\orbittwo{4}{4}=\frac{E_{7\left( 7\right) }(\fld{R})}{SO\left( 6,5;\fld{R}\right) \rtimes \fld{R}^{32}\times \fld{R}}.}
\end{array}
\end{equation}
Indeed, at the level of the semi-simple part of the orbit stabilizers, it holds that
\begin{equation}
\begin{array}{c}
SO(5,5; \fld{R})  \maxsymmsubsetneq SO\left( 6,5;\fld{R}\right) ; \\
SO(5,5; \fld{R}) \times SO\left( 1,1;\fld{R}\right)\maxsymmsubsetneq E_{6(6)}(\fld{R}).
\end{array}
\end{equation}
At the level of corresponding moduli spaces, \eref{eq:1-6} implies that \cite{Cerchiai:2010xv}
\begin{equation}
\modulitwo{2}{5}=M_{D=6}\rtimes\fld{R}^{16}=\frac{SO(5,5; \fld{R}) }{SO\left( 5,\fld{R}\right) \times SO\left( 5,\fld{R}\right) }\rtimes\fld{R}^{16}
\end{equation}
satisfies (the first embedding being trivial)
\begin{equation}\label{eq:1/2-embs}\fl
\modulitwo{2}{5}\subsetneq
\left\{\begin{array}{l}
\begin{array}{rl}
\displaystyle\modulitwo{2}{4}&=\mathcal{M}_{\mathrm{non\mbox{-}BPS, }D=4}\rtimes\fld{R}^{27}\\
&=M_{D=5}\rtimes\fld{R}^{27}=\displaystyle\frac{E_{6(6)}(\fld{R})}{USp\left( 8,\fld{R}\right) }\rtimes\fld{R}^{27};
\end{array}\\
\displaystyle\modulitwo{4}{4},
\end{array}\right.
\end{equation}
yielding the following relation:
\begin{equation}
\modulitwo{2}{5}\subset \left[ M_{D=5}\cap \modulitwo{4}{4}\right] .
\end{equation}
Furthermore, Eqs. \eref{eq:1/4-embs} and \eref{eq:1/2-embs} yield:
\begin{equation}
\modulitwo{4}{5}\maxsymmsubsetneq \modulitwo{2}{5}\subsetneq \modulitwo{4}{4}.
\end{equation}

Thus, it follows that a \emphsmall 1/2-BPS black hole, as well as a \emphsmall 1/4-BPS black hole, of $\SUSY=8$, $D=4$ supergravity can be uplifted to a \emphsmall 1/2-BPS black hole of $\SUSY=8$, $D=5$ supergravity.

\subsection{\texorpdfstring{U-duality orbits of $E_{7(7)}(\rng{Z})$}{U-duality orbits of E\_7(7)(Z)}}\label{sec:D4disc}

For quantized charges the continuous  U-duality is broken to an infinite discrete subgroup, which for $D=4$ is given by $E_{7(7)}(\rng{Z})\subset E_{7(7)}(\fld{R})$ \cite{Hull:1994ys}. The integral Freudenthal triple system $\mathfrak{M}(\mathfrak{J}^{\fld{O}_{\rng{Z}}^s}_{3})$ provides a natural model for $E_{7(7)}(\rng{Z})$, which may used to analyze the discrete U-duality orbits \cite{Krutelevich:2004}. The quantized black hole charge vector is given by,
\begin{equation}
x=\left(\begin{array}{cc}-q_0&P\\Q&p^0\end{array}\right), \ \  \mathrm{where}
\ \  q_0, p^0 \in \rng{Z} \ \  \mathrm{and} \ \  Q,
P\in\mathfrak{J}^{\fld{O}^{s}_{\rng{Z}}}_{3}.
\end{equation}
Note, the quartic norm and, hence, the entropy squared are quantized. In fact, $\Delta(x)$ is equal to either $4n$ or $4n+1$ for some $n\in\rng{Z}$. The discrete group $E_{7(7)}(\rng{Z})$ is defined by the set of invertible $\rng{Z}$-linear transformations $\sigma:\mathfrak{M}(\mathfrak{J}^{\fld{O}_{\rng{Z}}^s}_{3})\to\mathfrak{M}(\mathfrak{J}^{\fld{O}_{\rng{Z}}^s}_{3})$ preserving both the antisymmetric bilinear form and the quartic norm invariant,
\begin{equation}
\begin{array}{r@{\ =\ }l}
\{\sigma(x),\sigma(y)\}&\{x,y\},\\
\Delta(\sigma(x),\sigma(y),\sigma(z),\sigma(w))&\Delta(x,y,z,w).
\end{array}
\end{equation}

Like the previous examples in $D=5,6$, the charge conditions defining the orbits in the continuous theory are manifestly invariant under the discrete subgroup $E_{7(7)}(\rng{Z})$ and, hence, those states unrelated by U-duality in the classical theory remain unrelated in the quantum theory.   There are five disjoint \emph{classes} of orbits  corresponding to the five continuous orbits. Three of which are the small 1/2-BPS,  1/4-BPS and  1/8-BPS classes, with vanishing $I_4(x)$. There are two large classes of orbits, one 1/8-BPS and one non-BPS as determined by the sign of $I_4(x)$.   However, each of these classes is broken up into a countably infinite set of discrete orbits. To classify these orbits Krutelevich used $E_{7(7)}(\rng{Z})$ to bring an arbitrary charge vector into a \emph{diagonal reduced} canonical form. See \cite{Krutelevich:2004} and \ref{sec:fts}. However, unlike the previous case this canonical form is \emph{not} uniquely defined.

A partial classification of the orbits is achieved via the  set of four discrete invariants,
\begin{equation}\label{eq:4dinvs}
\begin{array}{r@{\ :=\ }l}
d_1(x)&\gcd(x),\\
d_2(x)&\gcd(3T(x,x,y)+x\{x,y\}) \, \forall y,\\
d_3(x)&\gcd(T(x,x,x)),\\
d_4(x)&\Delta(x).\\
\end{array}
\end{equation}
Note, there are two further arithmetic invariants appearing in the literature, but are not used here, see \ref{sec:intFTS}.

\paragraph*{$D=4$ diagonal reduced canonical form (Krutelevich 2004
\cite{Krutelevich:2004}):} Every element $x\in\mjz$ is
$E_{7(7)}(\rng{Z})$ equivalent to a diagonally reduced canonical form,
\begin{equation}\label{eq:4dcan}
x_{\mathrm{can}}=\alpha \left(\begin{array}{cc}  1 &  k\opname{diag}(1,l,lm) \\ 0 &
j \end{array}\right), \ \  \mathrm{where} \ \  \alpha>0.
\end{equation}
Note, $k\opname{diag}(1,l,lm)$ is the $D=5$ diagonally reduced canonical form \eref{eq:5dcan}. From here-on-in we will often use $(a,b,c)$ to mean $\opname{diag}(a,b,c)$.

While the $D=4$ canonical form for a generic charge vector is not uniquely determined by the discrete invariants \eref{eq:4dinvs} it is  uniquely specified for the subclass of black holes preserving more than 1/8 of the supersymmetries, i.e. rank 1 and rank 2 charge vectors \cite{Krutelevich:2004}. In this case the canonical form is simplified.

\paragraph*{$D=4$ Rank $<3$ diagonal reduced canonical form (Krutelevich 2004 \cite{Krutelevich:2004}):} Every element $x\in\mjz$ rank $<3$ and so preserving more than 1/8 of the supersymmetries is $E_{7(7)}(\rng{Z})$ equivalent to a diagonally reduced canonical form, \begin{equation}\label{eq:4dcansimplified}
\alpha\left(\begin{array}{cc}1&k(1,0,0)\\0&0\end{array}\right), \ \  \mathrm{where} \ \  \alpha >0.
\end{equation}
The simplified canonical form is uniquely determined by the two well defined arithmetic invariants from \eref{eq:4dinvs}, since
\begin{equation}
\begin{array}{r@{\ =\ }l}
d_1(Q_{\mathrm{can}})&\alpha,\\
d_2(Q_{\mathrm{can}})&2\alpha^{2}k,
\end{array}
\end{equation}
so that for arbitrary rank 1 or 2 $x$ one obtains $\alpha=d_1(x)$ and $k=(\sqrt{2}\alpha)^{-2}d_2(x)$. This facilitates the orbit classification for such states as is described below.

\paragraph*{$D=4$ Rank $<3$ black hole orbit classification (Krutelevich 2004 \cite{Krutelevich:2004}):}
\begin{enumerate}
\item The complete set of distinct 1/2-BPS charge vector orbits is given by,
\begin{equation}
\big\{\left(\begin{array}{cc}\alpha&0\\0&0\end{array}\right), \ \  \mathrm{where}
\ \  \alpha>0 \big\}.
\end{equation}

\item The complete set of distinct 1/4-BPS charge vector orbits is given by,
    \begin{equation}
\big\{\alpha\left(\begin{array}{cc}1&k(1,0,0)\\0&0\end{array}\right), \ \
\mathrm{where} \ \  \alpha, k>0 \big\}.
\end{equation}
\end{enumerate}

\subsection{Projective black holes}\label{sec:proj}

For black holes preserving less than 1/4 of the supersymmetries the analysis becomes increasing complex and the orbit classification for generic charge vectors is not known. However, for a subclass of such black holes, satisfying particular arithmetic conditions, the orbit classification is known. These black holes are referred to as \emph{projective}.

A black hole charge vector $x$ is said to be projective if its U-duality orbit contains a diagonal reduced element \eref{eq:4dcan} satisfying \cite{Krutelevich:2004, Borsten:2009zy},
\begin{equation}
\begin{array}{r@{\ =\ }l}
\gcd(\alpha k, \alpha j, (\alpha kl)^2 m) & 1;\\
\gcd(\alpha kl, \alpha j,(\alpha k)^2 lm) & 1;\\
\gcd(\alpha klm, \alpha j, (\alpha k)^2 l) & 1.
\end{array}
\end{equation}
One immediately sees that projectivity implies $\alpha=1$ in the canonical form  \eref{eq:4dcan} and therefore $\gcd(x)=1$. Black holes satisfying $\gcd(x)=1$ are conventionally referred to as \emph{primitive}.

While the general treatment of orbits in $ D = 4$  is lacking, the orbit representatives of projective black holes have been fully classified in \cite{Krutelevich:2004, Borsten:2009zy}.

\paragraph*{$D=4$ Projective black hole orbit classification (Krutelevich 2004 \cite{Krutelevich:2004}):} Any projective black hole charge vector $x$ is U-duality equivalent to an element,
\begin{equation}\label{eq:projcan}
\left(\begin{array}{cc}1& (1, 1,m)\\0& j\end{array}\right), \ \  \mathrm{where}\ \  j \in \{0, 1\}.
\end{equation}
The values of $m$ and $j$ are uniquely determined by $I_4(x)$. Further,
\begin{itemize}
\item $E_{7(7)}(\rng{Z})$ acts transitively on projective elements of a given norm $I_4(x)$.
\item If $I_4(x)$ is a squarefree\footnote{An integer is squarefree if its prime decomposition contains no repetition.} integer equal to 1 (mod 4) or if $I_4(x) = 4n$, where $n$ is squarefree and equal to 2 or 3 (mod 4), then $x$ is projective and hence U-duality acts transitively.
\end{itemize}
\emph{In the projective case all black holes with the same quartic norm and hence lowest order entropy are U-duality related.}

As already emphasized the generic case of not necessarily projective black holes is not fully understood.

\section{\texorpdfstring{Remark on the uplift of 1/2-BPS $D=4$ ``small'' black holes}{Remark on the uplift of 1/2-BPS D=4 ``small'' black holes}}\label{sec:Remark}

Within the analysis of maximal supergravity in $D=4,5,6$ in the \emph{continuous} U-duality regime performed in \autoref{sec:4/5-Rels} and \autoref{sec:5/6-Rels}, 1/2-BPS $D=4$ \emphsmall black holes have been uplifted to 1/2-BPS $D=5$ \emphsmall black holes (\autoref{sec:1/2}), which in turn have been shown to uplift to 1/2-BPS \emphsmall dyonic black strings (\autoref{sec:p=1-1/2}) and 1/2-BPS \emphsmall black holes/black $2$-branes (\autoref{sec:p=0,2-1/2}) in $D=6$.

However, within present treatment 1/2-BPS $D=4$ \emphsmall black holes have not been related to \emphlarge solutions of Einstein Eqs. coupled to maximal local supersymmetry.

This is due to the fact that such a class of asymptotically flat solutions of $\SUSY=8$, $D=4$ supergravity admit an uplift to a \emphlarge solution only starting from maximal $\SUSY=2$, $D=7$ supergravity. See \emph{e.g.} \cite{Ferrara:2008xz} and Refs. therein.

On the other hand, as given by the general analysis of \cite{Ferrara:1997ci}, in order to have an asymptotically flat \emphlarge solution of $\SUSY=2$, $D=7$ supergravity, an \emph{intersecting} configuration of black branes must be considered. See also \cite{D'Auria:1999fa}.  Some of these configurations, with simple, factorized near-horizon geometries, have been considered in Sect. 7 of \cite{Ferrara:2008xz}.

By analysing the spectrum of asymptotically flat black $p$-branes allowed in $D$ space-time dimensions by the bound $p\leqslant D-4$ \cite {Gibbons:1993sv,Ferrara:1997ci}, the lowest-dimensional \emphlarge solution of maximal supergravity to which 1/2-BPS $D=4$ \emphsmall black holes can be uplifted should be provided by 1/2-BPS dyonic black $2$-branes in $\SUSY=2$, $D=8$ supergravity. See \emph{e.g.} Sect. 8 of \cite{Ferrara:2008xz} and Refs. therein. The following one should then be given by 1/2-BPS dyonic black $3$-branes ($D3$-branes) of Type IIB maximal supergravity in $D=10$. See also the treatment given in Sect. 9 of \cite{Ferrara:2008xz}. We leave the interesting issue of this uplifts for future investigation.

It is here also worth remarking that, since $\SUSY=8$, $D=4$ supergravity might be expected to be all-loop UV finite \cite{Bern:2006kd,Bern:2009kd}, \emphsmall solutions within such a theory might not be expected to have their horizon stretched by quantum corrections, simply because these latter might not be there. Thus, such solutions would correspond to \emphsmall states in the spectrum of the theory at the full quantum level\footnote{Consistence issues related to such states have been recently addressed in \cite{Bianchi:2009wj,Bianchi:2009mj}.}. However, the same does not hold for (maximal) supergravity theories in $D\neq 4$. In particular, as mentioned in \cite{Bianchi:2009mj}, the regular solutions in $D=10$ and $D=11$ are expected to receive corrections at the quantum level, because a consistent UV completion of supergravities in such space-time dimensions would be provided by superstrings and $M$-theory, respectively.

\section{Conclusion}

We have summarized our current understanding of the black hole/string charge vector orbits  under the discrete U-dualities of $\SUSY=8$ supergravity in six, five and four dimensions. The discrete orbits of both the black strings in $D=6$ and the black holes/strings in $D=5$ \cite{Krutelevich:2002} admit a complete classification. Two distinct technical elements made this analysis tractable. First, the discrete U-duality groups, $SO(5,5;\rng{Z})$ in $D=6$ and $E_{6(6)}(\rng{Z})$  in $D=5$, may be modelled, in the sense of \cite{Gross:1996}, by the integral exceptional  quadratic and cubic Jordan algebras, respectively. These explicit representations, which both fundamentally rely  upon the ring of integral split-octonions, yielded diagonally reduced canonical forms for the charge vectors, from which the orbit representatives could, in principle, be obtained. Second, a complete list of independent arithmetic invariants, typically given by the gcd of irreps built out of the basic charge vector representations,   is known. These invariants are sufficient to uniquely fix the canonical form for a given charge vector. These two features together allow for the complete classification of the discrete orbits.

\begin{itemize}
\item $D=6$: the black string charge vector $\mathcal{Q}\in\jzs$ is $SO(5,5;\rng{Z}):=\opname{Str}_0(\jzs)$ equivalent to  a two charge diagonally reduced canonical form,
\begin{equation}\mathcal{Q}_{\mathrm{can}}=\left(\begin{array}{cc}k&0\\0&kl\end{array}\right),
\ \  k>0, \end{equation} which is uniquely determined by the two following
arithmetic invariants,
\begin{equation}\begin{array}{r@{\ :=\ }l}
b_1(\mathcal{Q})&\gcd(\mathcal{Q}),\\
b_2(\mathcal{Q})&N_2(\mathcal{Q})=\det \mathcal{Q}.
\end{array}
\end{equation}
\item $D=5$: the black hole (string) charge vector $Q\in\jz$ is $E_{6(6)}(\rng{Z}):=\opname{Str}_0(\jz)$ equivalent to  a three charge diagonally reduced canonical form,
\begin{equation}Q_{\mathrm{can}}=\left(\begin{array}{ccc}k&0&0\\0&kl&0\\0&0&klm\end{array}\right),
\ \  k>0,l\geq0, \end{equation} which is uniquely determined by the three
following arithmetic invariants,
\begin{equation}\begin{array}{r@{\ :=\ }l}
c_1(Q)&\gcd(Q),\\
c_2(Q)&\gcd(Q)^\sharp,\\
c_3(Q)&N_3(Q).
\end{array}
\end{equation}
\end{itemize}

The analogous treatment of the 4-dimensional black hole is not so transparent.  The integral FTS does indeed provide an elegant and natural representation of  the discrete U-duality group $E_{7(7)}(\rng{Z})$,  which again yields a diagonally reduced canonical charge vector. However, this canonical form is not uniquely determined by the known set of arithmetic U-duality invariants. The complete classification is known for two subcases: 1) Black holes preserving more than 1/8 of the supersymmetries 2) Black holes satisfying the projectivity condition.

\begin{itemize}
\item $D=4$ $>1/8$-BPS: the black hole charge vector $x\in\mj$ is $E_{7(7)}(\rng{Z}):=\opname{Aut}(\mjz)$ equivalent to  a two charge diagonally reduced canonical form,
\begin{equation}x_{\mathrm{can}}=\alpha\left(\begin{array}{cc}1&(k,0,0)\\0&0\end{array}\right),
\ \  \alpha>0,
\end{equation}
which is uniquely determined by the two following
arithmetic invariants,
\begin{equation}\begin{array}{r@{\ :=\ }l}
d_1(x)&\gcd(x),\\
d_2(x)&\gcd(3T(x,x,y)+\{x,y\}x), \ \  \forall y\in\mj.
\end{array}
\end{equation}
\item $D=4$ projective: the black hole charge vector $x\in\mj$ is $E_{7(7)}(\rng{Z}):=\opname{Aut}(\mjz)$ equivalent to  a four or five charge diagonally reduced canonical form,
\begin{equation}x_{\mathrm{proj can}}=\left(\begin{array}{cc}1& (1, 1,m)\\0& j\end{array}\right),
\ \  \mathrm{where}\ \  j \in \{0, 1\}.
\end{equation}
The values of $m$ and $j$ are uniquely determined by the quartic
$E_{7(7)}(\fld{R})$ invariant,  $I_4(x)$.
\end{itemize}
The orbit structure for generic 1/8-BPS and non-BPS charge vectors is far more complex. For example, it was shown in \cite{Krutelevich:2004} that the sub-example given by the FTS defined over the cubic Jordan algebra of split-complex $3\times 3$ Hermitian matrices is equivalent to the $SL(6, \rng{Z})$-orbits in $\wedge^3(\rng{Z}^6)$. This is an example that appeared in \cite{Bhargava:2004}, in which it was shown to be equivalent to the structure of balanced triples of ideal classes in quadratic rings. It would be interesting to understand what role, if any, these essentially number-theoretic observations might play in the physics of stringy black holes.

Evidently, there are a number of open questions. Chiefly, is it possible that the full space of 4-dimensional orbits could be resolved if the complete list of independent arithmetic invariants was known? For example, we have thus far used gcd of the $\rep{133}$ appearing in $\rep{56}\times_s\rep{56}$. What about the $\rep{1463}$?  Following \cite{Borsten:2008,Borsten:2008wd}, we may truncate to the 8 charges of the $STU$ model \cite{Sen:1995ff,Duff:1995sm,Gregori:1999ns,Bellucci:2008sv}, which transform as a $(\rep{2,2,2})$ of $SL(2,\rng{Z})\times SL(2,\rng{Z})\times SL(2,\rng{Z})$. Using this truncation, the $\rep{1463}$ in $\rep{56}\times_s \rep{56}$ reduces to the $\rep{(3,3,3)}$ in $(\rep{2,2,2})\times_s (\rep{2,2,2})$. Computing the gcd of this $\rep{(3,3,3)}$  gives the square of $d_1(x_{\mathrm{can}})$ and, therefore, adds no additional information. To proceed further, it would serve us well to have a full classification of the independent $E_{7(7)}(\rng{Z})$ arithmetic invariants.

It is tempting to extend this analysis to the various supergravity theories not considered here, but none-the-less have a Jordan algebraic  underpinning.  In particular, one might wish to consider the series of $\SUSY=2$ ``magic'' supergravities in $D=5$ and $D=4$, which have at their basis the cubic Jordan algebras defined over the four division algebras and the corresponding Freudenthal triple systems, respectively. However, this analysis is perhaps less well motivated, from both a physical and mathematical perspective. Physically, while the maximally supersymmetric theories considered here are expected to be protected from quantum corrections, these arguments do not in general hold for the less than maximal theories and, hence, the classical  U-dualities would typically be destroyed by quantum anomalies. Mathematically, it is known that not every element of the Jordan algebra of integral \emph{octonionic} $3\times 3$ Hermitian matrices, corresponding to the exceptional magic supergravities, is diagonalizable \cite{Elkies:1996}. Hence, the diagonally reduced canonical form used in the \emph{split-octonionic} $\SUSY=8$ case will not generalize to these exceptional magic supergravities. Finally, returning to the maximally supersymmetric theory in six dimensions, the black hole and membrane charges transform as the spinors \rep{16} and \rep{16'} of $SO(5,5;\fld{R})$, both of which may be represented as a pair of split-octonions \cite{Sudbery:1984}.  An integral structure could then be induced, as it was for the string, by using the integral split-octonions, again providing a natural framework with which to study the discrete U-duality orbits of $SO(5,5;\rng{Z})$.

\ack\addcontentsline{toc}{section}{Acknowledgments}

This work was supported in part by the STFC under rolling grant ST/G000743/1. L.B. would like to thank Dan Waldram for useful conversations. The work of S. Ferrara is supported by ERC Advanced Grant No. 226455, ``Supersymmetry, Quantum Gravity and Gauge Fields'' (SUPERFIELDS). A. M. would like to thank the Blackett Laboratory, Imperial College London, UK, where part of this work was done, for kind hospitality and stimulating environment. The work of A.M. has been supported by an INFN visiting Theoretical Fellowship at SITP, Stanford University, Stanford, CA, USA.

\appendix

\section{The integral split-octonions}\label{sec:octonions}

\subsection{Composition algebras}

An algebra $\mathds{A}$ defined over the reals $\fld{R}$ is said to be \emph{composition} if it has a non-degenerate quadratic form $|\bullet|:\mathds{A}\to\fld{R}$ such that for $a,b\in\mathds{A}$,
\begin{equation}
|ab|^2=|a|^2|b|^2,\ \  \forall a,b \in\alg,
\end{equation}
where we denote multiplicative product of the algebra by juxtaposition.

By considering $\fld{R}\subset\alg$ as the scalar multiples of the identity we may decompose $\mathds{A}$ into its real and imaginary parts $\alg=\fld{R}\oplus \alg'$, where $\alg'\subset\alg$ is the subspace orthogonal to $\fld{R}$. An arbitrary element $a\in\alg$ may be written $a=\Re(a) +\Im(a)$, where $\Re(a)\in\fld{R}$ and $\Im(a)\in \alg'$. The conjugation operation $a\mapsto \overline{a}$ defined by scalar multiplying all elements of $\alg'$ by $-1$ while leaving the all elements of $\fld{R}$ invariant satisfies \cite{Kantor:1980},
\begin{equation}
\overline{ab}=\overline{b}\overline{a},\qquad a\overline{a}=|a|^2.
\end{equation}
The natural  inner product defined by $2\langle a,
b\rangle=|a+b|^2-|a|^2-|b|^2$ is given, in terms of conjugation, by
\begin{equation}
\langle a, b\rangle=\Re(a\overline{b})=\Re(\overline{a}b).
\end{equation}

We denote respectively the commutator and associator by $[a,b]$ and $[a,b,c]$,
\begin{equation}
\begin{array}{r@{\ =\ }l}
[a,b]   & ab-ba, \\
\left[a,b,c\right] & \left(ab\right)c-a\left(bc\right).
\end{array}
\end{equation}
A composition algebra is said to be associative if the associator vanishes and commutative if the commutator vanishes. If the associator is an alternating function of its arguments then the algebra is said to be \emph{alternative}.

A \emph{division} algebra is a composition algebra satisfying the further requirement that it contain no zero-divisors,
\begin{equation*}
ab=0 \Rightarrow\ \  a=0\ \ \mathrm{or}\ \  b=0.
\end{equation*}
Hurwitz's celebrated theorem states that there are only four division algebras: the reals, complexes, quaternions and octonions denoted respectively by $\fld{R}, \fld{C}, \fld{H}$ and $\fld{O}$. These algebras are obtained by the Cayley-Dickson process. With a slight modification one can also generate their split signature cousins $\fld{C}^s, \fld{H}^s$ and $\fld{O}^s$, which are no longer division. We  assume familiarity with $\fld{R}, \fld{C}$ and  $\fld{H}$ and will only discuss here the most relevant case of the split-octonions, $\sO$. For a detailed account covering our many omissions the reader is referred to the   review by Baez \cite{Baez:2001dm}.

\subsection{The (integral) split-octonions}\label{sec:intoct}

The split-octonions are an 8-dimensional (non-division) composition algebra. They are both non-commutative and non-associative but are alternative. They may be generated from the split-quaternions via the Cayley-Dickson process. The split-quaternions are a 4-dimensional (non-division) composition algebra. The three imaginary units $i,j,k$ obey the following multiplication rules:
\begin{equation}
\begin{array}{rlrlrl}
i^2&=1,&j^2&=1,&k^2&=-1,\\
ij&=-ji=k,&ik&=-ki=j,&jk&=-kj=i.
\end{array}
\end{equation}
There is a convenient matrix representation of this algebra given
by,
\begin{equation}
\begin{array}{cc}
1=\left(\begin{array}{cc}1&0\\0&1\end{array}\right), &\qquad i=\left(\begin{array}{cc}1&0\\0&-1\end{array}\right),\\
j=\left(\begin{array}{cc}0&1\\1&0\end{array}\right), &\qquad
k=\left(\begin{array}{cc}0&1\\-1&0\end{array}\right),
\end{array}
\end{equation}
such that an arbitrary quaternion $a\in\fld{H}$ may be written,
\begin{equation}a=\left(\begin{array}{cc}a_{00}&a_{01}\\a_{10}&a_{11}\end{array}\right), \qquad
a_{ij}\in\fld{R}.
\end{equation}
The norm, real part and conjugation are given by,
\begin{equation}|a|^2=\det(a),\quad 2\Re(a) = \tr(a),\quad
\overline{a}=-ja^\mathrm{T}j,
\end{equation}
where $^\mathrm{T}$ denotes the matrix transpose.

The split-octonions $\sO$ may then be defined by introducing a fourth imaginary unit $\nu$,
\begin{equation}a+b\nu, \qquad a,b\in\fld{H}^s.
\end{equation}
The octonionic multiplication rules are defined as per the Cayley-Dickson process,
\begin{equation}(a+b\nu)(c+d\nu):=(ac-\overline{d}b)+(da+b\overline{c})\nu.
\end{equation}
The norm, real part and conjugation are given by,
\begin{equation}\label{eq:intoctfuncs}\fl\qquad
|a+b\nu|^2=\det(a)+\det(b),\quad
2\Re(a+b\nu) = \tr(a),\quad \overline{a+b\nu}=\overline{a}-b\nu.
\end{equation}

The ring of integral split-quaternions $\sQz$ is defined as the ring of $2\times 2$ matrices with entries in $\rng{Z}$. The norm and trace have integral values and the ring is closed under conjugation.

The ring of integral split-octonions $\sOz$ is then defined in the obvious manner and we may write an arbitrary integral split-octonions as,
\begin{equation}a+b\nu, \qquad a,b\in\sQz.
\end{equation}
The norm, the trace and conjugation, as defined in \eref{eq:intoctfuncs}, are well defined functions taking their values in $\rng{Z}$ and, moreover, $\sOz$ is a maximal order \cite{Weissman:2006,Coxeter:1946}.

\section{Jordan algebras}\label{sec:J}

A Jordan algebra $\mathfrak{J}$ is vector space defined over a ground field $\mathds{F}$ (we assume char$\mathds{F}\not=2,3$ throughout) equipped with a bilinear product satisfying,
\begin{equation}\label{eq:Jid}\fl\qquad
X\circ Y=Y\circ X, \qquad X^2\circ (X\circ Y)=X\circ (X^2\circ Y),
\qquad\forall\ X, Y \in \mathfrak{J}.
\end{equation}
While originally introduced with a view to generalizing the axiomatic basis of quantum mechanics \cite{Jordan:1933a, Jordan:1933b, Jordan:1933vh} Jordan algebras are now largely studied in their own right and have connections to numerous branches of mathematics, in particular exceptional Lie algebras, a fact we exploit here. For a detailed exposition of Jordan algebras and their historical development the reader is referred to \cite{Jacobson:1968, McCrimmon:2004}.

Using the nomenclature of \cite{McCrimmon:2004} the  subset of Jordan algebras relevant to supergravity may be  divided into three types: spin factors, quadratic factors and cubic factors. All three occupy important positions in various supergravity theories \cite{Kugo:1982bn,Gunaydin:1983bi,Gunaydin:1983rk}. In particular, the quadratic factors are relevant to the $D=6, \SUSY=8$ theory and the cubic factors are relevant to the $D=5, \SUSY=8$ theory and, accordingly, we focus  on these two examples here.

While the formal description of these algebras is not strictly necessary for our purposes, we could just as well dive straight in with the relevant explicit examples, we give a brief description in the following for completeness and, more importantly, as it facilitates the definition of the FTS required for the black holes in $D=4$.

\section{Jordan algebras and 6D black strings}\label{sec:6DJ}

\subsection{Quadratic Jordan algebras}\label{sec:quadJ}

A \emph{quadratic} form\footnote{We avoid using the conventional notation $Q$ for the quadratic form due to the plethora of $Q$'s representing electric charges.} $N_2$ on a vector space $V$ defined over a field $\mathds{F}$ is a homogeneous mapping from $V$ to $\mathds{F}$ of degree 2,
\begin{equation}N_2: V \to \mathds{F}\quad\mathrm{s.t.}\quad N_2(\alpha X)=\alpha^2 N_2(X)\quad \forall \alpha\in\mathds{F}, X\in V,
\end{equation}
such that its linearization,
\begin{equation}N_2(X,Y):=N_2(X+Y)-N_2(X)-N_2(Y)
\end{equation}
is bilinear.
A \emph{base point} is then defined as an element $c\in V$ satisfying $N_2(c)=1$. Given a space equipped with a quadratic form and possessing a base point we can define the \emph{trace} form,
\begin{equation}\Tr(X):=N_2(X,c).
\end{equation}
A quadratic Jordan algebra $\mathfrak{J}_2$ may derived from such a space by setting the identity $\mathds{1}=c$ and defining the Jordan product as,
\begin{equation}\label{eq:J2prod}
X\circ Y:=\frac{1}{2}(\Tr(X)Y+\Tr(Y)X- N_2(X,Y)\mathds{1}).
\end{equation}
On setting $X=Y$ one obtains
\begin{equation}\label{eq:deg2}
X^2-\Tr(X)X+N_2(X)\mathds{1}=0,\qquad\forall X\in\mathfrak{J}_2
\end{equation}
and $\mathfrak{J}_2$ is said to be of \emph{degree 2} \cite{McCrimmon:2004}. Moreover, on taking the trace of \eref{eq:deg2} one finds,
\begin{equation}N_2(X)=\frac{1}{2}[\Tr(X)^2-\Tr(X^2)],
\end{equation}
which is suggestively the form of the determinant of a $2\times 2$ matrix written in terms of the trace of powers and  powers of the trace.

There are three groups of particular importance associated with such quadratic Jordan algebras:
\begin{enumerate}
\item The \emph{automorphism} group $\opname{Aut}(\mathfrak{J}_{2})$ defined by the set of  invertible $\mathds{F}$-linear transformations $\sigma$ preserving the Jordan product,
\begin{equation}\sigma(X\circ Y)=\sigma(X)\circ\sigma(Y).
\end{equation}
The corresponding Lie algebra is given by the set of \emph{derivations} $\mathfrak{der}(\mathfrak{J}_{2})$,
\begin{equation}D(X\circ Y)=D(X)\circ Y +X\circ D(Y), \qquad \forall D\in \mathfrak{der}(\mathfrak{J}_{2}).
\end{equation}
\item The \emph{structure} group $\opname{Str}(\mathfrak{J}_{2})$ defined by the set of invertible $\mathds{F}$-linear transformations $\sigma$ preserving the quadratic norm up to a scalar factor,
\begin{equation}N_2(\sigma(X))=\alpha N_2(X), \qquad \alpha \in \mathds{F}.
\end{equation}
The corresponding Lie algebra $\mathfrak{Str}(\mathfrak{J}_{2})$ is given by,
\begin{equation}\mathfrak{Str}(\mathfrak{J}_{2})=L(\mathfrak{J}_{2})\oplus \mathfrak{der}(\mathfrak{J}_{2}),
\end{equation}
where $L(\mathfrak{J}_{2})$ denotes the set of left Jordan products $L_X(Y)=X\circ Y$.
\item The \emph{ reduced structure} group $\opname{Str}_0(\mathfrak{J}_{2})$ defined by the set of invertible $\mathds{F}$-linear transformations $\sigma$ preserving the quadratic norm,
\begin{equation}N_2(\sigma(X))= N_2(X).
\end{equation}
The corresponding Lie algebra $\mathfrak{Str}_0(\mathfrak{J}_{2})$ is given by factoring out scalar multiples of the identity in $L(\mathfrak{J}_{2})$,
\begin{equation}\mathfrak{Str}_0(\mathfrak{J}_{2})=L'(\mathfrak{J}_{2})\oplus \mathfrak{der}(\mathfrak{J}_{2}),
\end{equation}
where $L'(\mathfrak{J}_{2})$ denotes the set of left Jordan products by traceless elements, $L_X(Y)=X\circ Y$ where $\Tr(X) =0$.
\end{enumerate}
A $\opname{Str}_0(\mathfrak{J}_2)$ invariant rank may be assigned to elements in $\mathfrak{J}_2$ as in \autoref{tab:jordanrank2}.
\begin{table}[ht]
\caption{$\mathfrak{J}_2$ ranks.\label{tab:jordanrank2}}
\begin{tabular*}{\textwidth}{@{\extracolsep{\fill}}*{7}{M{c}}}
\toprule
& \multirow{2}{*}{Rank} && \multicolumn{2}{c}{Condition} & \\
\cmidrule(r){3-6}
&                       && X     & N_2(X) & \\
\midrule
& 0                     && =0    & =0     & \\
& 1                     && \neq0 & =0     & \\
& 2                     && \neq0 & \neq0\\
\bottomrule
\end{tabular*}
\end{table}

\subsection{\texorpdfstring{The Jordan algebra of split-octonionic $2\times 2$ Hermitian matrices and black strings}{The Jordan algebra of split-octonionic 2 x 2 Hermitian matrices and black strings}}\label{sec:22J}

Let us now focus our attention on the specific example relevant to the dyonic black strings of $D=6, \SUSY=8$ supergravity. We denote by $\mathfrak{J}^{\mathds{A}}_{2}$ the Jordan algebra of $2\times 2$ Hermitian matrices with entries in a composition algebra $\alg$ defined over the field $\mathds{F}$. We will assume $\mathds{F}=\fld{R}$ here.

An arbitrary element may be written as,
\begin{equation}X=\left(\begin{array}{cc}\alpha&a\\\overline{a}&\beta\end{array}\right), \ \ \mathrm{where}\ \  \alpha, \beta \in \fld{R}\ \ \mathrm{and}\ \  a\in\alg.
\end{equation}
The Jordan product \eref{eq:J2prod} is given by
\begin{equation}X\circ Y=\frac{1}{2}(XY+YX),\qquad X, Y \in \mathfrak{J}^{\mathds{A}}_{2},
\end{equation}
where juxtaposition denotes the conventional matrix product. The quadratic norm is simply given by the determinant,
\begin{equation}\label{eq:quadnorm}
N_2(X)=\det X=\alpha\beta-|a|^2.
\end{equation}

The reduced structure group is $SO(\dim \mathds{A}+1,1;\fld{R})$ for $\mathds{A}$ one of the division algebras $\fld{R}, \fld{C}, \fld{H}$ or $\fld{O}$ \cite{Sudbery:1984,Barton:2003}. However, setting $\mathds{A}=\sO$ the reduced structure becomes $SO(5,5; \fld{R})$, the $D=6, \SUSY=8$ U-duality group for real valued charges. An element of $\mathfrak{J}^{\sO}_{2}$ transforms as the vector \rep{10} of $SO(5,5;\fld{R})$. The quadratic norm in this case is nothing but the  quadratic singlet of $SO(5,5;\fld{R})$ in \rep{10\times 10} which we denote as $I_2$.

The $5+5$ electric/magnetic $D=6$ black string charges form a \rep{10} of $SO(5,5;\fld{R})$. Denoting them as $\mathcal{Q}_{r}$ ($r=1,...,10$ throughout)  the quadratic invariant may be written as
\begin{equation}\eta^{rs}\mathcal{Q}_{r}\mathcal{Q}_{s},
\end{equation}
where $\eta^{rs}$ is the $SO(5,5;\fld{R})$ metric,
\begin{equation}\eta=\left(\begin{array}{cc}0&\mathds{1}\\ \mathds{1}&0\end{array}\right).
\end{equation}

We may associate $\mathcal{Q}_{r}=(p^v, q_v)$, $v=0,\ldots4$, to an element $\mathcal{Q}\in\jzs$ in the following way,
\begin{equation}\mathcal{Q}=\left(\begin{array}{cc}p^0&Q_v\\\overline{Q}_v&q_0\end{array}\right),
\ \ \mathrm{where}\ \  q_0, p^0  \in \fld{R}\ \ \mathrm{and}\ \
Q_v\in\sO,
\end{equation}
and,
\begin{equation}\label{eq:dictD6}\fl\qquad
\begin{array}{rl}
Q_v&=\frac{1}{2}[(p^1+q_1)e_0+(p^2+q_2)e_1+(p^3+q_3)e_2+(p^4+q_4)e_3\\
&\phantom{=\frac{1}{2}[}+(p^1-q_1)e_4+(p^2-q_2)e_5+(p^3-q_3)e_6+(p^4-q_4)e_7],
\end{array}
\end{equation}
so that,
\begin{equation}N_2(\mathcal{Q})=\det \mathcal{Q}=I_2(\mathcal{Q}).
\end{equation}
The leading-order black string entropy is given by
\begin{equation}S_{D=6,\mathrm{BS}}\sim |I_2(\mathcal{Q})|= |N_2(\mathcal{Q})|.
\end{equation}

\subsection{\texorpdfstring{Integral Jordan algebra and black strings in $D=6, \SUSY=8$}{Integral Jordan algebra and black strings in D=6, N=8}}\label{sec:22Jz}

The corresponding integral Jordan algebra $\jzs$ is defined as the set of  $2\times 2$ Hermitian matrices defined over the ring of integral split-octonions defined in \ref{sec:intoct}. An arbitrary element may be written as,
\begin{equation}X=\left(\begin{array}{cc}\alpha&a\\\overline{a}&\beta\end{array}\right), \ \ \mathrm{where}\ \  \alpha, \beta \in \rng{Z}\ \ \mathrm{and}\ \  a\in\sOz.
\end{equation}

Evidently, $\jzs$ is not a linear Jordan algebra as it is not closed under the Jordan product.  It is, however, a well defined quadratic Jordan algebra \cite{McCrimmon:2004}.  Crucially, the quadratic norm and trace form take values in $\rng{Z}$.

The group $SO(5,5;\rng{Z})$ is defined as the set of invertible $\rng{Z}$-linear transformations leaving the quadratic norm invariant. Under the action of $SO(5,5;\rng{Z})$ every element of $\jzs$ is related to a diagonal reduced canonical form,
\begin{equation}\label{eq:6dcanform}
X_{\mathrm{can}}=\left(\begin{array}{cc}k&0\\0&kl\end{array}\right), \ \  k>0.
\end{equation}
It is not difficult to verify that that the set of invertible $\rng{Z}$-linear transformations $\sigma_{st}^{b}$ given by,
\begin{equation}
\sigma_{st}^{b}(X)=(\mathds{1}+bE_{st})X(\mathds{1}+\overline{b}E_{ts}),
\end{equation}
where $b\in\fld{O}_{\rng{Z}}^s$ and $E_{st}$ is a $2\times 2$ matrix with a single non-zero unit entry in the $st$ position, leave $N_2(X)$ invariant and so belong to $SO(5,5;\rng{Z})$. Explicitly, the action of $\sigma_{st}^{b}$ on $X$ is given by,
\begin{equation}
\begin{array}{r@{\ =\ }l}
\sigma_{12}^{b}(X)&\left(\begin{array}{cc}\alpha+\tr(b\overline{a})+\beta|b|^2&a+\beta a\\\overline{a}+\beta\overline{b}&\beta\end{array}\right),\\
\sigma_{21}^{b}(X)&\left(\begin{array}{cc}\alpha&a+\alpha\overline{b}\\\overline{a}+\alpha a&\alpha+\tr(b\overline{a})+|b|^2\end{array}\right).
\end{array}
\end{equation}
Through the successive application of these transformations a generic $X$ may be put into canonical form \eref{eq:6dcanform} by suitably modifying the iterative procedure presented in \cite{Krutelevich:2002}.

For an element $X$ of an integral Jordan algebra, an integer $d$ \emph{divides} $X$, denoted $d|X$, if $X=dX'$ with $X'$ integral. By taking the gcd of the the rank conditions \autoref{tab:jordanrank2} we may define the following set of arithmetic $SO(5,5;\rng{Z})$ invariants,
\begin{equation}
\begin{array}{r@{\ :=\ }l}
b_1(X)&\gcd(X),\\
b_2(X)&N_2(X).
\end{array}
\end{equation}
These are sufficient to fix the canonical form uniquely. Note, $b_1$, unlike $b_2$, is not an invariant of $SO(5,5;\fld{R})$.

\section{Jordan algebras and 5D black holes}\label{sec:5DJ}

\subsection{Cubic Jordan algebras}\label{sec:cubicJ}

There is a general prescription for constructing cubic Jordan algebras, due to Freudenthal, Springer and Tits \cite{Springer:1962, McCrimmon:1969, McCrimmon:2004}, for which all the properties of the Jordan algebra are essentially determined by the cubic form.  We sketch this construction here, following closely the conventions of \cite{Krutelevich:2004, McCrimmon:2004}.

Let $V$ be a vector space equipped with a cubic norm, i.e. a homogeneous map of degree 3
\begin{equation}\label{eq:cubicnorm}
N_3:V\to \mathds{F}, \ \ \mathrm{s.t.} \ \  N_3(\alpha X)=\lambda^3N_3(X), \ \  \forall \alpha \in \mathds{F}, X\in V
\end{equation}
such that its linearization,
\begin{equation}\fl
\begin{array}{rl}
N_3(X, Y, Z)&:=N_3(X+ Y+ Z)-N_3(X+Y)-N_3(X+ Z)-N_3(Y+Z)\\
&\phantom{:=}+N_3(X)+N_3(Y)+N_3(Z)
\end{array}
\end{equation}
is trilinear. If $V$ further contains a base point $N_3(c)=1, c\in V$  one may define the following four maps,
\begin{subequations}\label{eq:cubicdefs}
\begin{enumerate}
\item The trace,
    \begin{equation}
    \Tr(X)=N_3(c, c, X),
    \end{equation}
\item A quadratic map,
    \begin{equation}
    S(X)=N_3(X, X, c),
    \end{equation}
\item A bilinear map,
    \begin{equation}
    S(X, Y)=N_3(X, Y, c),
    \end{equation}
\item A trace bilinear form,
    \begin{equation}\label{eq:tracebilinearform}
    \Tr(X, Y)=\Tr(X)\Tr(Y)-S(X, Y).
    \end{equation}
\end{enumerate}
\end{subequations}
A cubic Jordan algebra $\mathfrak{J}$ with multiplicative identity $\mathds{1}=c$ may be derived from any such vector space if $N_3$ is \emph{Jordan cubic}, that is:
\begin{enumerate}
\item The trace bilinear form \eref{eq:tracebilinearform} is non-degenerate.
\item The quadratic adjoint map, $\sharp\colon\mathfrak{J}\to\mathfrak{J}$, uniquely defined by $\Tr(X^\sharp, Y) = N(X, X, Y)$, satisfies
    \begin{equation}\label{eq:Jcubic}
    (X^{\sharp})^\sharp=N_3(X)X, \qquad \forall X\in \mathfrak{J}.
    \end{equation}
\end{enumerate}
The Jordan product is then defined using,
\begin{equation}\label{eq:J3prod}
X\circ Y = \half\big(X\times Y+\Tr(X)Y+\Tr(Y)X-S(X,
Y)\mathds{1}\big),
\end{equation}
where, $X\times Y$ is the linearization of the quadratic adjoint,
\begin{equation}\label{eq:FreuProduct}
X\times Y = (X+Y)^\sharp-X^\sharp-Y^\sharp.
\end{equation}
Finally, the Jordan triple product is defined as
\begin{equation}\label{eq:Jtripleproduct}
\{X,Y,Z\}=(X\circ Y)\circ Z + X\circ (Y\circ Z)-(X\circ Z)\circ Y.
\end{equation}

There are three groups of particular importance associated with cubic Jordan algebras:
\begin{enumerate}
\item The \emph{automorphism} group $\opname{Aut}(\mathfrak{J}_{3})$ defined by the set of invertible $\mathds{F}$-linear transformations $\sigma$ preserving the Jordan product,
    \begin{equation}    \sigma(X\circ Y)=\sigma(X)\circ\sigma(Y).
    \end{equation}
    The corresponding
    Lie algebra is given by the set of \emph{derivations} $\mathfrak{der}(\mathfrak{J}_{3})$,
    \begin{equation}    D(X\circ Y)=D(X)\circ Y+X\circ D(Y), \qquad \forall D\in \mathfrak{der}(\mathfrak{J}_{3}).
    \end{equation}
\item The \emph{structure} group $\opname{Str}(\mathfrak{J}_{3})$ defined by the set of invertible $\mathds{F}$-linear transformations $\sigma$ preserving the quadratic norm up to a scalar factor,
    \begin{equation}    N_3(\sigma(X))=\alpha N_3(X), \qquad \alpha \in \mathds{F}.
    \end{equation}
    The corresponding Lie algebra $\mathfrak{Str}(\mathfrak{J}_{3})$ is
    given by,
    \begin{equation}    \mathfrak{Str}(\mathfrak{J}_{3})=L(\mathfrak{J}_{3})\oplus
    \mathfrak{der}(\mathfrak{J}_{3}), \end{equation} where $L(\mathfrak{J}_{3})$
    denotes the set of left Jordan products $L_X(Y)=X\circ Y$.
\item The \emph{ reduced structure} group $\opname{Str}_0(\mathfrak{J}_{3})$ defined by the set of invertible $\mathds{F}$-linear transformations $\sigma$ preserving the quadratic norm,
    \begin{equation}    N_3(\sigma(X))= N_3(X).
    \end{equation}
    The corresponding Lie algebra $\mathfrak{Str}_0(\mathfrak{J}_{3})$ is given by factoring out scalar multiples of the identity in $L(\mathfrak{J}_{3})$,
    \begin{equation}    \mathfrak{Str}_0(\mathfrak{J}_{3})=L'(\mathfrak{J}_{3})\oplus\mathfrak{der}(\mathfrak{J}_{3}),
    \end{equation}
    where $L'(\mathfrak{J}_{3})$ denotes the set of left Jordan products by traceless elements, $L_X(Y)=X\circ Y$ where $\Tr(X) =0$.
\end{enumerate}
A $\opname{Str}_0(\mathfrak{J}_3)$ invariant rank may be assigned to elements in $\mathfrak{J}_3$ as in \autoref{tab:jordanrank}.
\begin{table}[ht]
\caption[$\mathfrak{J}_3$ ranks]{$\mathfrak{J}_3$ ranks.\label{tab:jordanrank}}
\begin{tabular*}{\textwidth}{@{\extracolsep{\fill}}*{8}{M{c}}}
\toprule
& \multirow{2}{*}{Rank} && \multicolumn{3}{c}{Condition} & \\
\cmidrule(r){3-7}
&                       && X     & X^\sharp & N_3(X)     & \\
\midrule
& 0                     && =0    & =0       & =0         & \\
& 1                     && \neq0 & =0       & =0         & \\
& 2                     && \neq0 & \neq0\   & =0         & \\
& 3                     && \neq0 & \neq0\   & \neq0      &\\
\bottomrule
\end{tabular*}
\end{table}

\subsection{\texorpdfstring{The Jordan algebra of split-octonionic $3\times 3$ Hermitian matrices and black holes (strings)}{The Jordan algebra of split-octonionic 3 x 3 Hermitian matrices and black holes (strings)}}\label{sec:33J}

Let us now focus our attention on the specific example relevant to the black holes (strings) of $D=5, \SUSY=8$ supergravity. We denote by $\mathfrak{J}^{\mathds{A}}_{3}$ the cubic Jordan algebra of $3\times 3$ Hermitian matrices with entries in a composition algebra $\alg$ defined over the field $\mathds{F}$. We will assume $\mathds{F}=\fld{R}$ here.

An arbitrary element may be written as,
\begin{equation}\label{eq:cubicnormJ33}
X=\left(\begin{array}{ccc}\alpha&c&\overline{b}\\\overline{c}&\beta&a\\
b&\overline{a}&\gamma\end{array}\right), \ \ \mathrm{where}\ \ \alpha,
\beta, \gamma \in \fld{R}\ \ \mathrm{and}\ \  a, b, c\in\alg.
\end{equation}
The Jordan product \eref{eq:J3prod} is given by
\begin{equation}X\circ Y=\frac{1}{2}(XY+YX),\qquad X, Y \in \mathfrak{J}^{\mathds{A}}_{3},
\end{equation}
where juxtaposition denotes the conventional matrix product. The cubic norm \eref{eq:cubicnorm} is given by the determinant like object,
\begin{equation}\label{eq:cubicnormexp}
N_3(X)=\alpha\beta\gamma-\alpha a\overline{a}-\beta b\overline{b}-\gamma c \overline{c} +(ab)c+\overline{c}(\overline{b}\overline{c}). \end{equation}
The trace bilinear form \eref{eq:tracebilinearform} is given the conventional matrix trace,
\begin{equation}\label{eq:tracebilinearexp}
\Tr(X, Y)=\tr(X\circ Y).
\end{equation}
The quadratic adjoint \eref{eq:Jcubic} is given by,
\begin{equation}\label{eq:quadadjexp}
X^\sharp=\left(\begin{array}{ccc}\beta\gamma-|a|^2&\overline{b}\overline{a}-\gamma
c&ca-\beta \overline{b}\\ab-\gamma
\overline{c}&\alpha\gamma-|b|^2&\overline{c}\overline{b}-\alpha a\\
\overline{a}\overline{c}-\beta
b&bz-\alpha\overline{a}&\beta\alpha-|c|^2\end{array}\right).
\end{equation}
The elements $X\in \mathfrak{J}_{3}^{\mathds{A}}$ transform as the $(3\dim\mathds{A}+3)$ dimensional representation of the reduced structure group, $\opname{Str}_0({\mathfrak{J}_3^{\mathds{A}}})$. For $\mathds{A=R, C, H, O}$ $X\in\mathfrak{J}_{3}^{\mathds{A}}$ transforms as the \textbf{6, 9, 15, 27} of $SL(3,\fld{R})$, $SL(3,\fld{C})$, $SU^{*}(6)$, $E_{6(-26)}(\fld{R})$, respectively. These are the symmetries of the magic $\SUSY=2,D=5$ supergravities \cite{Gunaydin:1984ak,Gunaydin:1983bi,Gunaydin:1983rk} and the electric black hole charges fall into the corresponding representations.

Setting $\mathds{A}=\sO$ the reduced structure group becomes $E_{6(6)}(\fld{R})$, the $D=5, \SUSY=8$ U-duality group for real valued charges. Elements of $\mathfrak{J}^{\sO}_{3}$ transform as the fundamental \rep{27} of $E_{6(6)}(\fld{R})$. The cubic norm in this case is nothing but the  cubic singlet of $E_{6(6)}(\fld{R})$ in $\rep{27\times 27\times 27}$ which we denote as $I_3$. The quadratic adjoint \eref{eq:FreuProduct} gives the contragradient representation \rep{27'} in $\rep{27}\times_s \rep{27}$ (or equally the \rep{27} in $\rep{27'}\times_s \rep{27'}$). The trace bilinear form \eref{eq:tracebilinearform} gives the singlet in \rep{27\times 27'}.

We may associate the 27 electric black hole charges to an element $Q\in\J$ in the following way,
\begin{equation}\fl\qquad
Q=\left(\begin{array}{ccc}
q_1            & Q_v            & \overline{Q}_s \\
\overline{Q}_v & q_2            & Q_c            \\
Q_s            & \overline{Q}_c & q_3
\end{array}\right), \ \ \mathrm{where}\ \  q_1, q_2, q_3  \in \fld{R}\ \ \mathrm{and}\ \  Q_{v, s, c} \in\sO,
\end{equation}
so that,
\begin{equation}N_3(Q)=I_3(Q).\end{equation}
The leading-order black hole entropy is given by
\begin{equation}
S_{D=5, \mathrm{BH}}=
\pi\sqrt{|I_3(Q)|}=\pi\sqrt{ |N_3(Q)|}.
\end{equation}

Similarly, we may associate the 27 magnetic black string charges to an element $P\in\J$ in the following way,
\begin{equation}\fl\qquad
P=\left(\begin{array}{ccc}
p_1            & P_v            & \overline{P}_s \\
\overline{P}_v & p_2            & P_c            \\
P_s            & \overline{P}_c & p_3
\end{array}\right), \ \ \mathrm{where}\ \  p_1, p_2, p_3  \in \fld{R}\ \ \mathrm{and}\ \  P_{v, s, c} \in\sO,
\end{equation}
so that,
\begin{equation}N_3(P)=I_3(P).\end{equation}
The leading-order black string entropy is given by
\begin{equation}S_{D=5, \mathrm{BS}}= \pi\sqrt{|I_3(P)|}=\pi\sqrt{|N_3(P)|}.\end{equation}

\subsection{\texorpdfstring{Integral cubic Jordan algebras and black holes (strings) in $D=5, \SUSY=8$}{Integral cubic Jordan algebras and black holes (strings) in D=5, N=8}}\label{sec:33Jz}

The corresponding integral Jordan algebra $\jz$ is defined as the set of  $3\times 3$ Hermitian matrices defined over the ring of integral split-octonions defined in \ref{sec:intoct} \cite{Krutelevich:2002}. An arbitrary element may be written as,
\begin{equation}X=\left(\begin{array}{ccc}\alpha&a&\overline{b}\\\overline{a}&\beta&c\\ b&\overline{c}&\gamma\end{array}\right), \ \ \mathrm{where}\ \  \alpha, \beta, \gamma \in \rng{Z}\ \ \mathrm{and}\ \  a, b, c\in\sOz.
\end{equation}
$\jz$ is not closed under the Jordan product, however, the cubic norm and trace bilinear form are integer valued, which are the crucial properties for our purposes. Moreover,  $\jz$  is closed under the quadratic adjoint map and its linearization as required.

The group $E_{6(6)}(\rng{Z})$ is defined as the set of invertible $\rng{Z}$-linear transformations leaving the cubic norm invariant. It was shown in \cite{Krutelevich:2002} that under the successive application of such discrete transformations every element of $\jz$ is related to a diagonal reduced canonical form,
\begin{equation}\label{eq:5dcanform}
X_{\mathrm{can}}=\left(\begin{array}{ccc}k&0&0\\0&kl&0\\0&0&klm\end{array}\right), \ \  k>0, l\geq 0.
\end{equation}
By taking the gcd of the the rank conditions \autoref{tab:jordanrank} we may define the following set of independent arithmetic $E_{6(6)}(\rng{Z})$ invariants,
\begin{equation}
\begin{array}{r@{\ :=\ }l}
c_1(X)&\gcd(X),\\
c_2(X)&\gcd(X^\sharp),\\
c_3(X)&N_3(X).
\end{array}
\end{equation}
These are sufficient to fix the canonical form uniquely. Note, $c_1$ and $c_2$, unlike $c_3$, are not  invariants of $E_{6(6)}(\fld{R})$.

\section{The Freudenthal triple system and 4D black holes}\label{sec:fts}

Given a cubic Jordan algebra $\mathfrak{J}_3$ defined over $\fld{R}$, there exists a corresponding  FTS given by the vector space $\mathfrak{M}(\mathfrak{J}_3)$,
\begin{equation}
\mathfrak{M}(\mathfrak{J}_3)=\mathds{R\oplus R}\oplus
\mathfrak{J}_3\oplus\mathfrak{J}_3.
\end{equation}
An arbitrary element $x\in \mathfrak{M}(\mathfrak{J}_3)$ may be written as a ``$2\times 2$ matrix'',
\begin{equation}
x=\left(\begin{array}{cc}\alpha&X\\Y&\beta\end{array}\right), \ \ \mathrm{where}
~\alpha, \beta\in\fld{R}\ \ \mathrm{and}\ \  X, Y\in\mathfrak{J}_3.
\end{equation}
For convenience we identify the quantity
\begin{equation}
\kappa(x):=\half(\alpha\beta-\Tr(X,Y)).
\end{equation}
The FTS comes equipped with a non-degenerate bilinear antisymmetric quadratic form, a quartic form and a trilinear triple product \cite{Freudenthal:1954,Brown:1969,Faulkner:1971, Ferrar:1972, Krutelevich:2004}:
\begin{subequations}
\begin{enumerate}
\item Quadratic form $ \{\bullet, \bullet\}$: $\mathfrak{M}(\mathfrak{J}_3)\times\mathfrak{M}(\mathfrak{J}_3)\to \fld{R}$
    \begin{equation}\label{eq:bilinearform}
    \begin{array}{c}
    \{x, y\}=\alpha\delta-\beta\gamma+\Tr(X,Z)-\Tr(Y,W),\\
    \mathrm{where\qquad}x=\left(\begin{array}{cc}\alpha&X\\Y&\beta\end{array}\right),\qquad y=\left(\begin{array}{cc}\gamma&W\\ Z&\delta\end{array}\right).
    \end{array}
    \end{equation}
\item Quartic form $\Delta:\mathfrak{M}(\mathfrak{J}_3)\to \fld{R}$
    \begin{equation}\label{eq:quarticnorm}
    \Delta (x)=-4[\kappa(x)^2+(\alpha N(X)+\beta N(Y)-\Tr(X^\sharp, Y^\sharp))].
    \end{equation}

\item Triple product $T:\mathfrak{M}(\mathfrak{J}_3)\times
\mathfrak{M}(\mathfrak{J}_3)\times\mathfrak{M}(\mathfrak{J}_3)\to\mathfrak{M}(\mathfrak{J}_3)$ which is uniquely defined by
\begin{equation}
\{T(x, y, w), z\}=2\Delta(x, y, w, z),
\end{equation}
where $\Delta(x, y, w, z)$ is the full linearization of $\Delta(x)$ normalized such that $\Delta(x, x, x, x)=\Delta(x)$. The explicit form of $T(x)=T(x,x,x)$ is given:
\begin{equation}\label{eq:Tofx}\fl
\begin{array}{r@{\ =\ }l}
T(x)& \left(\begin{array}{cc}T_\alpha&T_X\\T_Y&T_\beta\end{array}\right)\\
& 2\left(\begin{array}{cc}-\alpha\kappa(x)-N(Y)&-(\beta Y^\sharp-Y\times X^\sharp)+\kappa(x)X\\
(\alpha X^\sharp-X\times Y^\sharp)-\kappa(x)Y&\beta\kappa(x)+N(X)\end{array}\right).
\end{array}
\end{equation}
\end{enumerate}
\end{subequations}
Note that all the necessary definitions, such as the cubic and trace bilinear forms, are inherited from the underlying Jordan algebra $\mathfrak{J}_3$.

The \emph{automorphism group} $\opname{Aut}(\mathfrak{M}(\mathfrak{J}_3))$ is given by the set of all invertible $\fld{R}$-linear transformations which leave both $\{x, y\}$ and $\Delta (x, y, w, z)$ invariant \cite{Brown:1969}. Note, for any transformation $\sigma\in \opname{Aut}(\mathfrak{M}(\mathfrak{J}_3))$ we have
\begin{equation}
T(\sigma(x), \sigma(y), \sigma(w))=\sigma(T(x, y, w)).
\end{equation}
The corresponding Lie algebra is given by \cite{Jacobson:1971},
\begin{equation}\mathfrak{Aut}(\mathfrak{M}(\mathfrak{J}_3))=\mathfrak{J}_3\oplus\mathfrak{J}_3\oplus\mathfrak{Str}(\mathfrak{J}_3).
\end{equation}

The Freudenthal triple systems,  defined over various Jordan algebras, and their associated automorphism groups are summarized in \autoref{tab:FTSsummary}. This table covers a number supergravities  of interest: $\SUSY=2$ $STU$, $\SUSY=2$ coupled to $n$ vector multiplets; magic $\SUSY=2$ and $\SUSY=8$.  The heterotic string with $\SUSY=4$ supersymmetry and $SL(2,\fld{R})\times SO(6,22; \fld{R})$ U-duality may also be included by using the Jordan algebra $\fld{R}\oplus Q_{5,21}$ \cite{Pioline:2006ni,Gunaydin:2009dq}.
\begin{table}
\caption[Jordan algebras, corresponding FTSs, and their associated symmetry groups]{The automorphism  group $\opname{Aut}(\mathfrak{M}(\mathfrak{J}_3))$ and the dimension of its representation $\dim\mathfrak{M}(\mathfrak{J}_3)$ given by the Freudenthal construction defined over the cubic Jordan algebra $\mathfrak{J}_3$ with dimension $\dim\mathfrak{J}_3$ and reduced structure group $\opname{Str}_0(\mathfrak{J}_3)$.\label{tab:FTSsummary}}
\begingroup
\footnotesize
\begin{tabular*}{\textwidth}{@{\extracolsep{\fill}}*{5}{M{c}}}
\toprule
\mathrm{Jordan\ algebra\ }\mathfrak{J}_3 & \opname{Str}_0(\mathfrak{J}_3) & \dim\mathfrak{J}_3 & \opname{Aut}(\mathfrak{M}(\mathfrak{J}_3)) & \dim\mathfrak{M}(\mathfrak{J}_3) \\
\midrule
\fld{R}                     & -                                  & 1   & SL(2,\fld{R})                                   & 4     \\
\fld{R}\oplus\fld{R}           & SO(1,1;\fld{R})                       & 2   & SL(2,\fld{R})\times SL(2,\fld{R})                  & 6    \\
\fld{R}\oplus\fld{R}\oplus\fld{R} & SO(1,1;\fld{R})\times SO(1,1;\fld{R})    & 3   & SL(2,\fld{R})^{\times3} & 8     \\
\fld{R}\oplus \Gamma_n           & SO(1,1;\fld{R})\times SO(n-1,1;\fld{R})   & n+1 & SL(2,\fld{R})\times SO(2,n;\fld{R})                & 2n+4  \\
J_{3}^{\fld{R}}             & SL(3,\fld{R})                         & 6   & Sp(6,\fld{R})                                   & 14    \\
J_{3}^{\fld{C}}             & SL(3,\fld{C})                         & 9   & SU(3,3;\fld{R})                                 & 20    \\
J_{3}^{\fld{H}}             & SU^\star(6,\fld{R})                   & 15  & SO^\star(12,\fld{R})                            & 32    \\
J_{3}^{\fld{O}}             & E_{6(-26)}(\fld{R})                   & 27  & E_{7(-25)}(\fld{R})                             & 56    \\
J_{3}^{\fld{O}^s}           & E_{6(6)}(\fld{R})                     & 27  & E_{7(7)}(\fld{R})                               & 56   \\
\bottomrule
\end{tabular*}
\endgroup
\end{table}

The conventional concept of matrix rank may be generalized to Freudenthal triple systems in a natural and $\opname{Aut}(\mathfrak{M}(\mathfrak{J}_3))$ invariant manner as in \autoref{tab:FTSrank} \cite{Ferrar:1972, Krutelevich:2004}.
\begin{table}[ht]
\caption[$\mathfrak{J}_3$ ranks]{$\mathfrak{M}(\mathfrak{J}_3)$ ranks.\label{tab:FTSrank}}
\begin{tabular*}{\textwidth}{@{\extracolsep{\fill}}*{9}{M{c}}}
\toprule
& \multirow{2}{*}{Rank} && \multicolumn{3}{c}{Condition} \\
\cmidrule(r){3-8}
&                       && x     & 3T(x,x,y)+\{x,y\},\;\forall\ y &T(x,x,x)       &\Delta(x) &\\
\midrule
& 0                     && =0    & =0                             & =0            & =0       &\\
& 1                     && \neq0 & =0                             & =0            & =0       &\\
& 2                     && \neq0 & \neq0\                         & =0            & =0       &\\
& 3                     && \neq0 & \neq0\                         & \neq0         & =0       &\\
& 4                     && \neq0 & \neq0\                         & \neq0         & \neq0\\
\bottomrule
\end{tabular*}
\end{table}

Let us now focus our attention on the specific example relevant to the black holes of $D=4, \SUSY=8$ supergravity. We denote by $\mj$ the FTS defined over the cubic Jordan algebra of $3\times 3$ Hermitian matrices with entries in split-octonions.  Elements of $\mj$ transform as the fundamental \rep{56} of $E_{7(7)}(\fld{R})$. The quartic norm in this case is nothing but the  unique quartic singlet of $E_{7(7)}(\fld{R})$ in $\mathbf{56 \times 56 \times 56\times56}$ which we denote as $I_4$. The triple product \eref{eq:Tofx} gives the fundamental  \rep{56} in $\mathbf{56\times 56\times56}$. The antisymmetric bilinear form \eref{eq:bilinearform} gives the singlet in $ \mathbf{56}\times_a \rep{56} $.

We may associate the $28+28$ electric/magnetic black hole charges to an element $x\in\mj$ in the following way,
\begin{equation}
x=\left(\begin{array}{cc}-q_0&P\\Q&p^0\end{array}\right), \ \  \mathrm{where}
\ \  q_0, p^0 \in \fld{R} \ \  \mathrm{and} \ \  Q,
P\in\mathfrak{J}^{\fld{O}^s}_{3},
\end{equation}
where $p^0, q_0$ are the graviphotons and $P,Q$ are the magnetic/electric $\mathbf{27}'$ and \rep{27} respectively, so that,
\begin{equation}\Delta(x)=I_4(x).
\end{equation}

The leading-order black hole entropy is given by
\begin{equation}\label{eq:4-entropy}
S_{}=\pi\sqrt{|I_4(x)|}=\pi\sqrt{|\Delta(x)|}.
\end{equation}

\subsection{The integral FTS and 4D black holes}\label{sec:intFTS}

To place an integral structure on the FTS we use, following \cite{Krutelevich:2004}, the integral Jordan algebra $\jz$ and define,
\begin{equation}\mjz:=\rng{Z}\oplus\rng{Z}\oplus\jz\oplus\jz.
\end{equation}
An arbitrary element may be written,
\begin{equation}
x=\left(\begin{array}{cc}\alpha&X\\Y&\beta\end{array}\right), \ \
\mathrm{where} \ \  \alpha, \beta \in \rng{Z} \ \  \mathrm{and}
\ \  X, Y\in\jz.
\end{equation}
The quartic norm and antisymmetric bilinear form are both integer valued and consequently $T(x,x,x)\in\mjz$.  The quartic norm $\Delta(x)$ is either $4n$ or $4n+1$ for some $n\in\rng{Z}$.

The discrete U-duality group $E_{7(7)}(\rng{Z})$ is defined as the set of invertible $\rng{Z}$-linear transformation preserving the quartic norm and antisymmetric bilinear form. It is generated by the following three maps \cite{Brown:1969, Krutelevich:2004}:
\begin{subequations}
\begin{eqnarray}\label{eq:FreudenthalConstructionTransformations}
\fl\quad\phi(W) &: \left(\begin{array}{cc}\alpha&X\\Y&\beta\end{array}\right) \mapsto \left(\begin{array}{cc}\alpha+(Y,W)+(X, W^{\sharp})+\beta N(W)&X+\beta W\\
                                                                                 Y+X\times W+\beta W^\sharp  &\beta
                                                                 \end{array}\right),\\
\fl\quad\psi(Z) &: \left(\begin{array}{cc}\alpha&X\\Y&\beta\end{array}\right) \mapsto \left(\begin{array}{cc}\alpha& X+Y\times Z+\alpha Z^\sharp\\
                                                                                Y+\alpha Z &\beta+(X,Z)+(Y, Z^{\sharp})+\alpha N(Z)
                                                                 \end{array}\right),\\
\fl\quad T(s) &: \left(\begin{array}{cc}\alpha&X\\Y&\beta\end{array}\right)    \mapsto \left(\begin{array}{cc}\alpha & s(X)\\
                                                                                {s'}^{-1}(Y) &\beta
                                                                 \end{array}\right).
\end{eqnarray}
\end{subequations}
where $s\in \opname{Str}_0(\jz)$  and $s'$ is its adjoint defined with respect to the trace bilinear form, $\Tr(s(X), s'(Y))=\Tr(X, Y)$.

Using these transformations it was shown in \cite{Krutelevich:2004} that every element $x\in\mjz$ is equivalent to a diagonally reduced canonical form,
\begin{equation}\label{eq:4dcan2}
x_{\mathrm{can}}=\alpha \left(\begin{array}{cc}  1 &  X_{\mathrm{can}} \\ 0
& j \end{array}\right), \ \  \mathrm{where} \ \  \alpha>0.
\end{equation}
Here $X_{\mathrm{can}}=k(1,l,lm)$ is the diagonally reduced canonical form of elements in $\jz$. However, the uniqueness of this canonical form is not guaranteed.

For an element $x$ of an integral FTS, an integer $d$ \emph{divides} $x$, denoted $d|x$, if $x=dx'$ with $x'$ integral. By taking the gcd of the the rank conditions \autoref{tab:FTSrank} we may define the following set of independent arithmetic $E_{7(7)}(\rng{Z})$ invariants,
\begin{equation}
\begin{array}{r@{\ :=\ }l}
d_1(x)&\gcd(x),\\
d_2(x)&\gcd(3T(x,x,y)+\{x,y\}x),\;\forall y\\
d_3(x)&\gcd(T(x,x,x)),\\
d_4(x)&\Delta(x).
\end{array}
\end{equation}
Note, $d_1, d_2$ and $d_3$, unlike $d_4$, are not  invariants of $E_{7(7)}(\fld{R})$. Additionally, we may also define,
\begin{equation}\label{eq:DiscreteInvariants}
\begin{array}{r@{\ :=\ }l}
d'_2(x)&\gcd(\mathcal{P}(x),\,\mathcal{Q}(x),\, \mathcal{R}(x))\\
d'_4(x)&\gcd(x \wedge T(x)),
\end{array}
\end{equation}
where $\wedge$ denotes the antisymmetric tensor product.
$\mathcal{P}(x)=Y^\sharp -\alpha X$ and $\mathcal{Q}(x)=X^\sharp
-\beta Y$ are the charge combinations appearing in the 4D/5D lift
\cite{Borsten:2009zy, Gaiotto:2005gf} and
$\mathcal{R}(x):\mathfrak{J}_3\to \mathfrak{J}_3$ is a Jordan
algebra endomorphism given by
\begin{equation}
\mathcal{R}(x)(Z)=2\kappa(x)Z+2\{X,Y,Z\},
\end{equation}
where $\{X,Y,Z\}$ is the Jordan triple product \eref{eq:Jtripleproduct}. Taken together, $(\mathcal{P}(x),\,\mathcal{Q}(x),\, \mathcal{R}(x))$ form the adjoint representation of  the 4-dimensional U-duality: $\rep{133}$ in the case of $E_{7(7)}(\rng{Z})$. Under the 5-dimensional U-duality, they transform as the fundamental, contragredient fundamental and adjoint representations, respectively: $\rep{27}$, $\rep{27'}$ and $\rep{1+78}$ in the case of  $E_{6(6)}(\rng{Z})$. Moreover, after subtracting the symplectic trace, $x \wedge T(x)$ transforms as the $\rep{1539}$ in $\rep{56}\times_a\rep{56}$.

Evaluated on the canonical form \eref{eq:4dcan2} one obtains,
\begin{equation}\label{eq:DiscreteInvariants_jklm}
\begin{array}{r@{\ =\ }l}
d_1(x_{\mathrm{can}})&\alpha\\
d_2(x_{\mathrm{can}})&\alpha^2\gcd(j,2k)\\
d'_2(x_{\mathrm{can}})&\alpha^2\gcd(j,k)\\
d_3(x_{\mathrm{can}})&\alpha^3\gcd(j,2k^2l)\\
d_4(x_{\mathrm{can}})&\alpha^4(j^2+4k^3l^2m)\\
d'_4(x_{\mathrm{can}})&\alpha^4\gcd(j,k^2l).
\end{array}
\end{equation}
Note, $d'_2(x_{\mathrm{can}})$ is refined compared to $d_2(x_{\mathrm{can}})$. Unlike the $D=5$ case the invariants \eref{eq:DiscreteInvariants_jklm} are insufficient to determine uniquely $j,k,l,m$, as can be seen by taking any example with $j=1$. Note, however, that $\alpha$ is clearly fixed by $d_1(x)$.  Consequently, the reduced canonical form \eref{eq:4dcan2} of any given black hole is not necessarily unique and, to the best of our knowledge, there is no complete classification of the U-duality orbits. For example,
\begin{equation}
\begin{array}{r@{\ =\ }l}
x &\alpha\left(\begin{array}{cc}1 & (0,0,0) \\ (0,0,0) & j\end{array}\right), \\
x'&\alpha\left(\begin{array}{cc}1 & (j,0,0) \\ (0,0,0) & j\end{array}\right),
\end{array}
\end{equation}
are both in canonical form and U-duality related using $\phi(W)$ in \eref{eq:FreudenthalConstructionTransformations} with $W=(1,0,0)$.

\subsection{Projective elements}\label{sec:appproj}

An element $x$ is said to be  \emph{projective} if its U-duality orbit contains a diagonal reduced element \cite{Krutelevich:2004}, \begin{equation}x= \left(\begin{array}{cc}  \alpha &  (X_1, X_2, X_3) \\ 0 & \beta \end{array}\right),
\end{equation}
satisfying
\begin{equation}\label{eq:projective element def}
\begin{array}{r@{\ =\ }l}
\gcd(\alpha X_1, \alpha\beta, X_2X_3)&1;  \\
\gcd(\alpha X_2, \alpha\beta, X_1X_3)&1;  \\
\gcd(\alpha X_3, \alpha\beta, X_1X_2)&1.  \\
\end{array}
\end{equation}
The concept of a projective element was originally introduced for the case $\mathfrak{J}_3=\rng{Z}\oplus\rng{Z}\oplus\rng{Z}$ along with certain generalizations central to the new view on Gauss composition and its extension as expounded in \cite{Bhargava:2004}.

The class of projective FTS elements is of particular relevance to recent developments in number theory \cite{Bhargava:2004, Krutelevich:2004}. Note,
\begin{itemize}
\item If $x$ is projective then $d'_2(x)=1$.
\item If $d_3(x)=1$ then $x$ is projective.
\item If $d_3(x) \geq 3$ or $T(x)=0$ then $x$ is not projective.
\item When $\Delta$ is odd, $d_3(x)=1$ iff $x$ is projective.
\end{itemize}
While the general treatment of the $E_{7(7)}(\rng{Z})$ orbits is lacking, the orbit representatives of  \emph{projective} elements have been fully classified in \cite{Krutelevich:2004}, at least for $\mathfrak{J}=\mathfrak{J}_{3}^{\mathds{A}}$ where $\mathds{A}$ is one of the three integral split  composition algebras $\fld{C}^s, \fld{H}^s$ or $\fld{O}^s$.  Any projective element $x$ is U-duality equivalent to an element \cite{Krutelevich:2004}:
\begin{equation}\label{eq:projective_fts}
\left(\begin{array}{cc}1 & (1,1,m) \\(0,0,0) &j \\ \end{array}\right),\ \ j\in \{0, 1\},\ m\in\rng{Z},
\end{equation}
where the values of $m$ and $j$ are uniquely determined by $\Delta(x)$.

\section*{References}
\addcontentsline{toc}{section}{References}

%

\end{document}